# Generative AI in Action: Field Experimental Evidence from Alibaba's Customer Service Operations

Xiao Ni[a*], Yiwei Wang[b*], Tianjun Feng[c], Lauren Xiaoyuan Lu[d],
Yitong Wang[e], and Congyi Zhou[e]

[a] Fudan University, School of Management, Shanghai, China, xni20@fudan.edu.cn

[b] Zhejiang University, International Business School, Zhejiang, China, yiweiwang@intl.zju.edu.cn

[c] Fudan University, School of Management, Shanghai, China, tfeng@fudan.edu.cn

[d] Dartmouth College, Tuck School of Business, Hanover, NH 03755, lauren.x.lu@tuck.dartmouth.edu

[e] Alibaba Group Inc.

[*] The first two authors contributed equally to this project.

July 2026

**Abstract**

In collaboration with Alibaba, we study how a generative AI assistant affects service performance in e-commerce after-sales operations. In a large-scale field experiment, human agents providing digital chat support were randomly assigned access to a gen AI assistant. The assistant drafts issue diagnoses and solution proposals in the opening stage only; agents can adopt, modify, or disregard them. Because of this discretion, we estimate the effects of both gen AI access and usage. On average, gen AI improves service speed and subjective service quality, measured by customer ratings, but has no significant effect on objective service quality, measured by customer retrials. These gains come from more than automation. Gen AI reshapes agent-customer interactions: treated agents respond faster and take a more proactive role, while customers provide less input; both patterns persist into later chat stages. These average effects, however, mask heterogeneity across agents. Lower-performing agents benefit the most, indicating that gen AI can narrow performance gaps. Top-performing agents experience declines in both subjective and objective service quality. This decline is consistent with workflow disruption: among top performers, gen AI use increases shift-away time, response delays, and immediate customer retrials, suggesting weakened service continuity in the focal chat. Successful gen AI deployment therefore requires careful performance evaluation and rollout tailored to agent skill.

## 1 Introduction

The rapid advancement of AI is changing how organizations perform and manage work. Earlier waves of AI adoption created operational value mainly by improving prediction, classification, routing, and other structured decisions. Yet the role of these tools in service settings remained limited because many service interactions require contextual understanding and interpersonal communication that earlier AI systems were less able to provide (Huang and Rust 2018). Generative AI, powered by sophisticated language models, expands the role of AI by assisting workers in tasks that combine information processing with natural-language communication. This shift is especially important for customer service operations, where performance depends not only on whether customer issues are resolved, but also on how service is delivered and experienced (Grönroos 1984, Parasuraman et al. 1985).

Online customer support provides a natural setting for studying this shift. In e-commerce, customer service is both high-volume and labor intensive: firms receive large numbers of time-sensitive inquiries, and human agents often handle several customer chats at the same time. The work also combines two goals that are difficult to separate. Agents must resolve order-related problems, such as refunds, exchanges, and delivery delays, while also shaping the customer experience through timely and appropriate communication. This combination of procedural problem solving and conversational service is precisely where gen AI may be useful. Unlike earlier AI tools that mainly classified requests or routed customers, gen AI assistants can generate context-specific text suggestions during the interaction while leaving the final decision to human agents. This design creates both an opportunity and a management challenge: gen AI may improve service performance, but its value depends on how agents incorporate AI into their work.

The rise of gen AI also connects to broader debates about the workforce implications of AI. One view is that gen AI can augment workers by automating routine tasks and giving less experienced employees access to better diagnostic and communication routines (Agrawal A. et al. 2023, Autor 2024). Another view is that reliance on AI may weaken independent judgment, disrupt established routines, or create new forms of workflow cost (Agrawal N. et al. 2023, Dell'Acqua 2022). These views imply different predictions about performance gaps across workers. If gen AI substitutes for missing experience, lower-performing workers may benefit the most. If it complements existing expertise, higher-performing workers may gain more. If it disrupts expert routines, however, high-performing workers may even be harmed.

Motivated by these tensions, we ask three questions: *whether gen AI improves service performance, how it changes agent-customer interactions, and whether its effects differ across workers with different baseline performance*.

We study these questions in collaboration with Alibaba through a large-scale randomized field experiment in its e-commerce after-sales service operations. Human agents providing digital chat support were randomly assigned access to a gen AI assistant. The assistant offered two core functions: *diagnosis* of customer issues and *solution proposals*. Both were delivered as text suggestions that agents could adopt, modify, or disregard. Because agents choose whether to use the suggestions, usage is neither complete nor random. We therefore estimate the intention-to-treat (ITT) effect of gen AI access and use random assignment as an instrument to estimate the local average treatment effect (LATE) of actual usage.

Our results show that gen AI improves *service speed* and *subjective service quality*. Treated agents identify customer issues more quickly, complete chats in less time, receive higher customer ratings, and have lower dissatisfaction rates. However, gen AI has no significant effect on *objective service quality*, measured by whether customers return with the same issue within three days. This divergence shows that gen AI can create value through service *interaction* and customer experience even when it does not improve the underlying *resolution* outcome. It also suggests that productivity measures based only on objective resolution may miss an important part of the value created by gen AI in service work.

The service performance gains come from more than automation. To examine how gen AI changes agent-customer interactions, we divide each chat into three stages: issue identification and solution proposal, resolution, and confirmation. Although the assistant is deployed primarily in the first stage, its effects extend beyond the moment of direct tool use. Assisted by gen AI, agents become more proactive in service: they respond faster and take a larger role in the exchange, both across the full chat and in later stages. This shift reduces the contribution required from customers, who respond with more short affirmations, upload fewer pictures, and send shorter, less lexically complex messages. Together, these patterns suggest that gen AI reshapes the service interaction rather than merely automating diagnosis and solution formulation.

Further analysis shows that the effects of gen AI are uneven across agents. When we segment agents by quintiles of pretreatment customer ratings, lower-performing agents achieve larger gains in subjective service quality, and reductions in overall chat duration are concentrated among lower-performing agents. This pattern suggests that gen AI can narrow performance gaps. In contrast, agents in the top performance quintile experience declines in both subjective and objective service quality, reflected in lower customer ratings, higher dissatisfaction rates, and higher retrial rates. Process evidence points to a workflow-disruption interpretation. In Alibaba's concurrent-chat environment, gen AI usage among top performers increases their later-stage shift-away time, slows their responses during resolution and confirmation, and raises customers' immediate retrials, all of which together suggest weakened service continuity in the focal chat. This pattern is unlikely to reflect overreliance on AI suggestions: top performers use and adhere to the gen AI assistant less often, and survey evidence indicates greater skepticism toward its outputs.

We make three contributions. The first is to research on AI and worker productivity. Existing field evidence shows that gen AI raises worker output in B2B customer service settings (Brynjolfsson et al. 2025). In our B2C e-commerce setting, we distinguish service interaction from service resolution. Estimating the causal effect of actual gen AI usage (LATE) and of access (ITT), we show that gen AI improves service speed and subjective service quality (customer ratings) without improving objective service quality (customer retrials). This interaction-versus-resolution divergence suggests that neither resolution-based nor perception-based metrics alone capture gen AI's effect on service performance.

The second is to research on behavioral dynamics in human–AI interaction. Prior field evidence attributes AI's effect on customer sentiment to the content of agents' messages, which become more empathetic, informative, and solution-oriented (Zhang and Narayandas 2026). Using stage-level process data from chat logs, we trace how the agent-customer interaction unfolds on both sides. Gen AI's effects extend beyond the stage in which it is deployed: agents respond faster and take a larger role in the exchange even in later stages where the gen AI assistant is no longer active. Alongside this shift, customers contribute less to the communication — they send more brief affirmations, fewer pictures, and shorter messages.

The third is to research on AI and the workforce. We find that gen AI compresses the performance distribution from both ends, as lower-performing agents gain in speed and quality while top performers' service quality declines. Process evidence is consistent with attention reallocation across concurrent chats rather than with broad overreliance on AI suggestions, indicating that in concurrent-chat service environments, gen AI can impose costs on top performers by disrupting their existing routines.

In terms of managerial implications, our findings suggest that successful gen AI deployment requires careful performance evaluation and smart deployment, not simply providing access to the technology. Gen AI can improve service speed and perceived service quality. However, because these gains do not necessarily translate into better objective resolution outcomes, managers should evaluate gen AI using a broad set of service measures, including variables related to both customer perceptions and actual problem resolution. The heterogeneous effects also caution against a uniform rollout. For lower-performing agents, gen AI may standardize service delivery and narrow performance gaps. For top-performing agents, managers should monitor workflow disruption and ensure that productivity targets do not compromise service continuity.

## 2 Literature Review

Our study contributes to three related streams of research on AI deployment in business operations: (1) AI and worker productivity; (2) behavioral dynamics in human-AI interaction; and (3) AI and the workforce. Together, these streams frame a central question in service operations: *where does gen AI create value when it enters human service work*? The productivity literature asks whether AI improves performance and which dimensions of performance may change. The human-AI interaction literature explains how workers incorporate algorithmic assistance into their routines. The workforce literature examines how those effects vary across workers and potential impact on employment. We connect these questions by examining service performance, worker-customer interaction, and heterogeneous impact on workers within the same field experiment.

### 2.1 AI and Worker Productivity

Early research in management and operations viewed AI mainly as an automation technology for structured tasks such as forecasting, scheduling, routing, allocation, and procurement (Van Donselaar et al. 2010, Cui et al. 2022, Sun et al. 2022, Boyaci et al. 2024). These systems create value by automating prediction or supporting decisions. Customer service is harder to automate because performance depends on both issue resolution and interpersonal communication (Luo et al. 2019, Schanke et al. 2021, Xu et al. 2026). Gen AI expands the scope of AI support by producing context-specific language during live service interactions.

Recent field studies show that AI assistance can improve service performance and productivity. Zhang and Narayandas (2026) find that AI-generated suggestions improve response speed and customer sentiment in online chats, with effects that vary by customer intent and prior chatbot experience. Brynjolfsson et al. (2025) show that a GPT-based assistant increases resolutions per hour in B2B technical support, with larger gains for less-skilled and less-experienced workers. Related studies of knowledge work reach similar conclusions but also identify important boundary conditions. GitHub Copilot helps software engineers complete coding tasks more quickly (Peng et al. 2023), and ChatGPT improves writing speed and quality, particularly for lower-skilled workers (Noy and Zhang 2023). Performance may decline, however, when workers use gen AI for tasks outside the technology's capability frontier (Dell'Acqua et al. 2026). This evidence establishes that gen AI can improve output, while suggesting that its value depends on the task, the performance measure, and the way the tool is incorporated into work.

Our study is closely related to Brynjolfsson et al. (2025), but it examines a different service process and a broader question about how gen AI creates value. Brynjolfsson et al. establish that gen AI raises labor productivity in B2B technical support, where interactions are relatively long and complex and performance is measured primarily by resolutions per hour. We study B2C e-commerce after-sales service, where agents handle short, high-volume, concurrent chats under relatively standardized resolution procedures. In this setting, gen AI may create value by changing service interaction even when it does not change service resolution. We therefore distinguish among service speed, subjective service quality (customer ratings), and objective service quality (customer retrials). Gen AI shortens issue identification time and chat duration, raises customer ratings, and lowers dissatisfaction, but it does not reduce three-day customer retrials. This interaction-versus-resolution divergence differs from the null average service quality effects in Brynjolfsson et al. and is consistent with gen AI having a stronger effect on the customer experience of service in our B2C setting than on the service resolution outcome.

The two studies also differ in what they reveal about the heterogeneous performance effects. Brynjolfsson et al. provide evidence consistent with an AI system disseminating the practices of more capable workers, thereby increasing the productivity of less skilled agents. We use granular, stage-level process data to trace how assistance in issue identification and solution proposal carries into later stages where the tool is no longer deployed. The results show that AI-assisted agents respond more quickly and contribute a larger share of messages, while customers provide less input. We also observe a distinct downside among top-performing agents: their service speed improves, but their subjective and objective service quality deteriorates. Longer shift-away time and slower responses are consistent with disruption to the service continuity of focal chats. Thus, Brynjolfsson et al. (2025) establish that gen AI can raise productivity in technical support and provide evidence consistent with the diffusion of effective practices. We complement their study by showing that gen AI can improve service interaction without improving underlying resolution outcomes, reshape agent-customer interactions beyond the stage of direct AI assistance, and create workflow disruption costs for top-performing agents.

### 2.2 Behavioral Dynamics in Human-AI Interaction

A second stream of research examines behavioral dynamics in human-AI interaction, especially overreliance, aversion, and discretionary use (Lebovitz et al. 2022). Access to an AI tool does not imply uniform use. Workers may accept, edit, ignore, or avoid algorithmic advice, and these choices can shape both average and heterogeneous treatment effects.

Research on overreliance shows that users may defer too much to algorithmic recommendations or reduce their own effort when responsibility is shared with technology (Karau and Williams 1993). Dell'Acqua (2022) finds that workers often reduce effort as AI tools improve, a pattern sometimes described as *falling asleep at the wheel*. Balakrishnan et al. (2026) and Snyder et al. (2026) show that individuals may place too much weight on algorithmic predictions, especially when systems appear accurate, thereby amplifying automation bias. At the same time, reliance on AI can improve performance when users combine algorithmic advice with their own experience (Bastani et al. 2026).

Research on AI aversion shows the opposite pattern: users may underuse AI, abandon algorithms after minor mistakes (Dietvorst et al. 2018), or override recommendations in favor of personal expertise (Waardenburg et al. 2022, Caro and de Tejada Cuenca 2023). Aversion can be reduced when users view tasks as objective (Castelo et al. 2019), understand algorithmic logic (Lehmann et al. 2022), or can adjust AI outputs (Dietvorst et al. 2018). Hou et al. (2024) similarly show that AI transparency and perceived smartness can increase adoption in online medical consultations.

Our study examines how gen AI changes the agent-customer interaction process, beyond whether agents accept or reject individual AI suggestions. Because not all treated agents use the tool, we use LATE estimates to distinguish the effect of access from the effect of actual use. We find that gen AI use changes agent-customer interactions even after the AI-assisted stage. Agents respond faster and send a larger share of the messages, both across the full chat and in later stages. Customers, in turn, send more short affirmative replies, upload fewer pictures, and write shorter and simpler messages. These results suggest that gen AI affects service work beyond automation by making agents more proactive while reducing customer contribution throughout the interaction.

### 2.3 AI and the Workforce

A third stream of research examines the workforce implications of AI. Earlier debates often focused on whether AI augments or displaces labor. More recent work asks how AI reorganizes work: which tasks are automated, which remain human-handled, which workers benefit, and how coordination changes when AI is inserted into existing workflows.

Several studies emphasize augmentation. Autor (2024) argues that gen AI can democratize expertise and expand the scope of middle-skill work. Acemoglu et al. (2022a) and McElheran et al. (2024) show that AI adoption in firms can enhance existing capabilities rather than replace workers outright. In operational settings, Bai et al. (2022) find that algorithmic work assignment can raise productivity when workers perceive the assignment process as fair. These studies suggest that AI can complement labor by improving access to information and reducing task frictions.

Other studies point to distributional risks. Zolas et al. (2021) show that AI-intensive job creation is concentrated in high-tech sectors, while Acemoglu et al. (2022b) find that AI adoption has not produced broad wage growth. Capraro et al. (2024) and Hosseini Maasoum and Lichtinger (2025) caution that gen AI may disproportionately benefit experienced workers when it complements existing expertise. Field evidence from customer support instead finds larger gains for less-experienced or lower-skilled workers, consistent with AI substituting for missing knowledge or communication routines (Brynjolfsson et al. 2025, Zhang and Narayandas 2026).

We extend this literature by showing that heterogeneity in the effects of gen AI can include losses among top performers, rather than merely smaller gains. Lower-performing agents achieve the largest improvements in service speed and subjective service quality, whereas top-performing agents experience declines in both subjective and objective service quality. The heterogeneity therefore extends beyond the smaller gains among higher-skilled agents documented by Brynjolfsson et al. (2025). Our process evidence is consistent with workflow disruption as a possible explanation. For top-performing agents, gen AI usage increases their shift-away time and response time, and customers' immediate retrials. These findings underscore that gen AI is not uniformly beneficial to workers. Its effects depend on how well the tool fits workers' established routines.

## 3 Experiment Setup

### 3.1 Company Background

In this study, we collaborated with Alibaba, one of the largest e-commerce companies in the world by gross merchandise value in fiscal year 2023.[1] The field experiment was conducted on Taobao, the flagship e-commerce platform of Alibaba. Similar to Amazon Marketplace, Taobao connects millions of merchants with consumers and handles a high volume of transactions. This environment provides an ideal setting for studying how gen AI affects online customer service operations. As of November 2023, 84% of official customer service agents worked entirely remotely, while the remaining agents worked from local offices. Agents worked independently, and the performance of one agent did not directly affect the performance of other agents.

### 3.2 Online After-Sales Service Chats

Our study focuses on online after-sales service chats conducted through the digital platforms of Alibaba. A service chat includes all text exchanges between a customer and a human agent within an interactive session. We focus on order-related after-sales issues, including exchanges, returns, refunds, repairs, and shipping-fee inquiries. These issues account for 56.5% of all customer service chats and constitute the most common reasons for customer contact. As of January 2024, Taobao had more than 380 million average daily active users and approximately 0.4 million daily requests for online after-sales chats during the experiment period.

To manage this volume, Taobao uses a two-tier service process. Customers first interact with a chatbot, which retrieves order histories and suggests solutions from a predefined set of actions. If the chatbot does not resolve the issue, the customer can be routed to a human agent within the same chat interface. Customer-agent matching is implemented through a greedy matching algorithm, which remained unchanged during the field experiment.[2] After a chat session ends, customers are invited to rate the service on a one-to-five scale, with higher values indicating better perceived service quality.

Human agents also have access to a set of standard support tools embedded in the sidebar of the chat interface. The sidebar contains two main components. The first is a customer information panel, which displays relevant details such as contact information, order history, and prior interactions. The second is a predefined response library, which agents can search by keyword to retrieve suggested replies and corresponding standard operating procedures for specific customer issues. These tools support consistency and efficiency in customer-agent communication.

---

[1] Alibaba 2023 Annual Report, https://static.alibabagroup.com/reports/fy2023/ar/ebook/en/index.html

[2] The matching algorithm uses agents' historical performance over a fixed evaluation period as one of its inputs. Agents in different pretreatment performance quintiles may therefore face different task environments, including differences not fully captured by broad issue categories. Treatment assignment was randomized within this existing matching system, and the matching rule remained unchanged throughout the experiment.

### 3.3 The Gen AI Assistant

In 2023, Alibaba developed a gen AI assistant to support human agents in online customer service chats. The assistant combines Qwen LLMs with machine learning algorithms fine-tuned for customer support scenarios on Taobao. The system was trained on service chats, order details, and customer information to support issue identification and response generation in after-sales service. Once deployed, the gen AI assistant provided agents with two types of text suggestions during customer chats:

1) A *diagnosis* to identify the issues associated with a customer order.
2) A *solution* to resolve the customer issues.

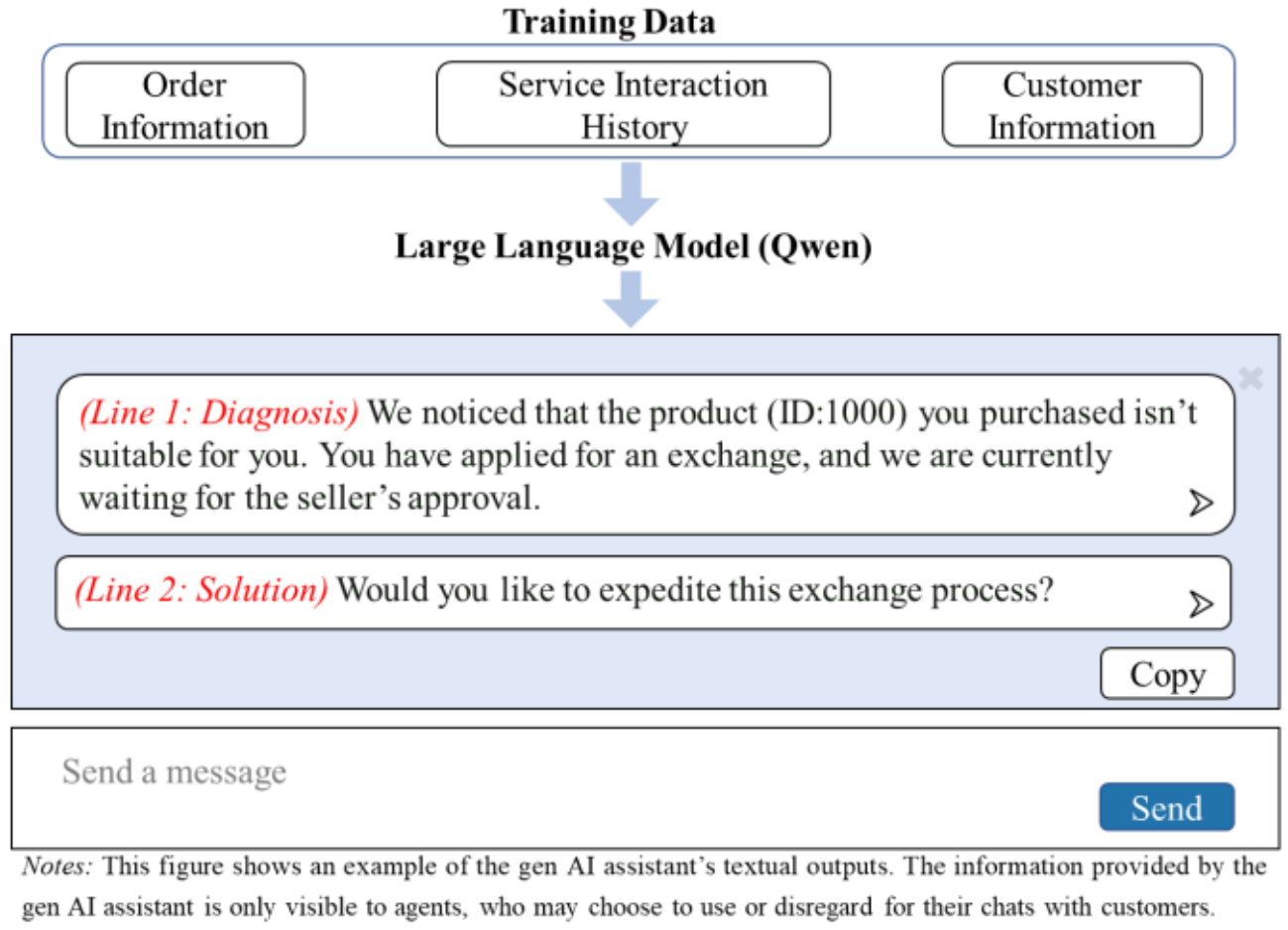

*Notes:* This figure shows an example of the gen AI assistant's textual outputs. The information provided by the gen AI assistant is only visible to agents, who may choose to use or disregard for their chats with customers.

**Figure 1. An Illustration of the Gen AI Assistant's Functions**

Figure 1 provides an example of the functions of the gen AI assistant. In this example, the assistant generates two lines of text. The first line identifies the issue, such as the status of an exchange request. The second line proposes a possible next step, such as expediting the exchange process. These suggestions are designed to help agents respond more promptly during the issue identification and solution proposal stage of the chat.

The AI-generated suggestions are visible only to agents. Agents can send a suggestion directly with one click, copy it into the chat box and revise it before sending, or disregard it and compose a response manually. This design preserves agent discretion and allows us to distinguish access to gen AI from actual usage of gen AI suggestions.

### 3.4 Experiment Design

Alibaba conducted a four-week randomized A/B test from January 23, 2024, to February 19, 2024, to evaluate the effect of the gen AI assistant on after-sales customer service performance. The experiment involved 5,940 new customer service agents with tenure of less than one year. During the experiment, these agents handled approximately 2.56 million chats and received approximately 0.39 million customer ratings.

Among the participating agents, 2,895 agents, or 48.7%, were randomly assigned to the treatment group, and 3,045 agents, or 51.3%, were randomly assigned to the control group. Randomization was conducted at the agent level based on agent ID. The experimental sample accounted for 15.6% of official customer service agents on Taobao at the time of the experiment.[3]

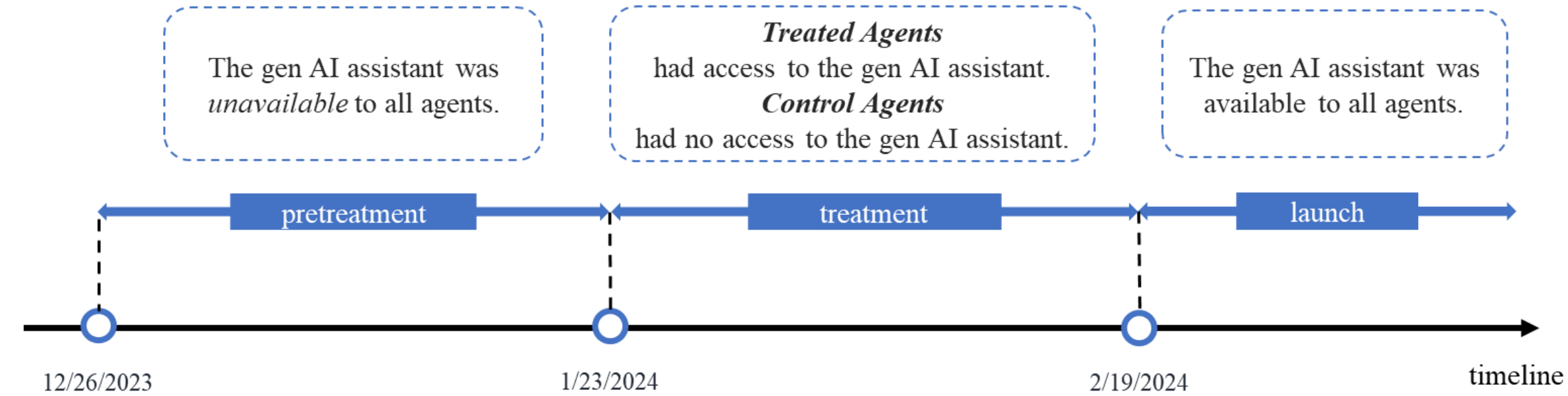

**Figure 2. Experiment Timeline**

Figure 2 summarizes the experiment timeline. The study period covers eight weeks, including a four-week pretreatment period and a four-week treatment period. From December 26, 2023, to January 22, 2024, the gen AI assistant was unavailable to all agents. From January 23, 2024, to February 19, 2024, treated agents had access to the gen AI assistant, while control agents continued to work without it. The field experiment ended on February 19, 2024, and the assistant was launched to all agents on February 20, 2024.

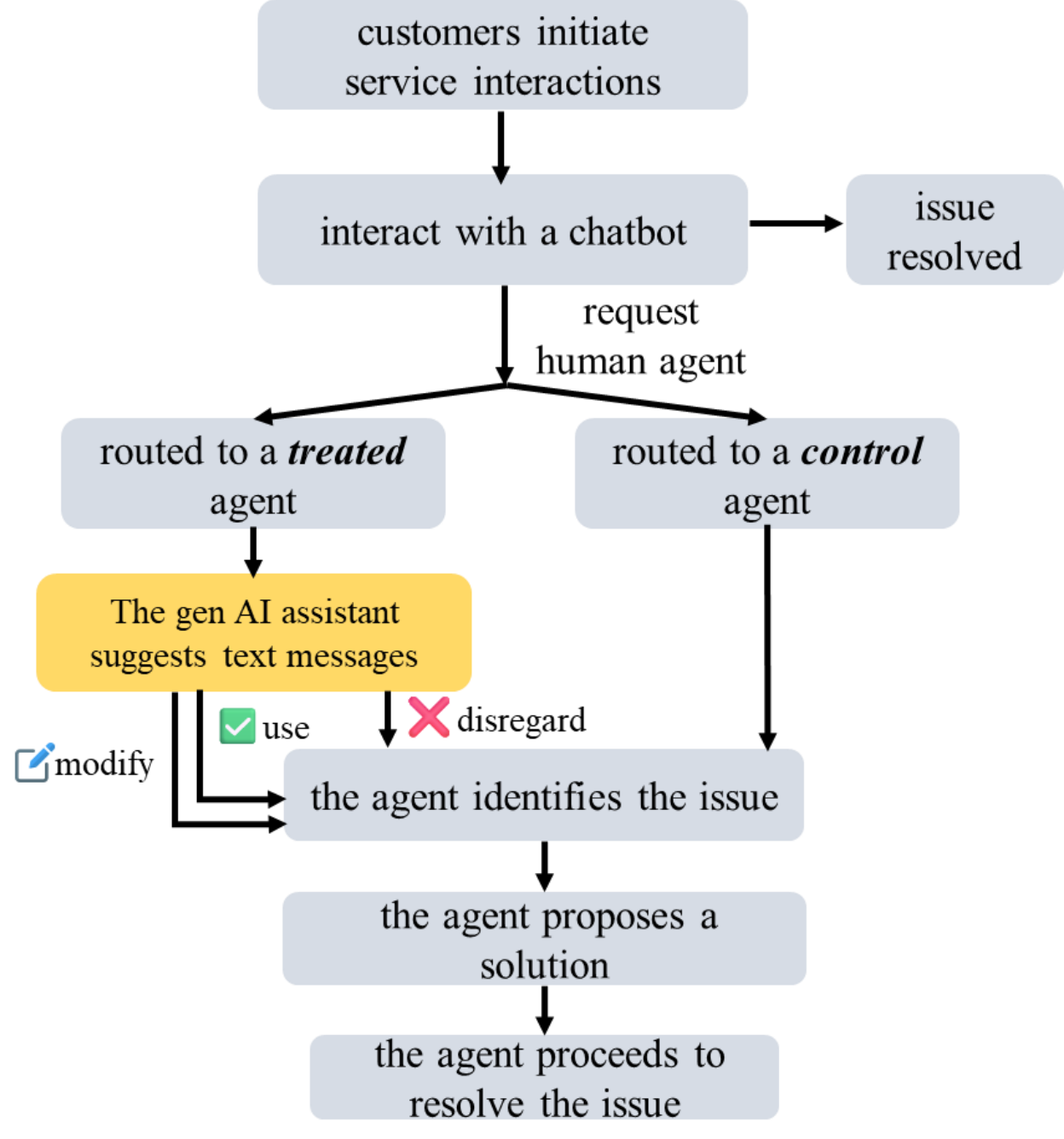

**Figure 3. Experiment Treatment Process**

[3] In this study, all experiment participants were official Taobao agents rather than the agents affiliated with individual sellers. Official Taobao agents are responsible for managing all after-sales customer issues for the entire platform, including complaints and disputes.

Figure 3 illustrates the treatment process. After interacting with the customer service chatbot, customers who requested human support were assigned to agents using the same matching algorithm used before the experiment. Control agents had access only to the standard support tools described in Section 3.2. Treated agents had access to the same standard tools plus an additional message box displaying suggestions generated by the gen AI assistant described in Section 3.3. Treated agents could adopt, modify, or disregard these suggestions.

### 3.5 Data Structure

Our dataset contains three main components: *agent information*, *session logs*, and *chat content*. Agent information includes treatment assignment, gender, age, and tenure. Session logs contain structured attributes of each service interaction, including chat ID, agent ID, customer ID, start and end times, concurrency, issue category, issue identification time, agent typing time, agent response time, customer rating, and follow-up contacts. Chat content includes message-level information, including sender identity, timestamps, message type, and message content. We merge these sources to construct a chat-level dataset with more than 2.56 million chats and 0.39 million customer ratings during the experiment. Table 1 reports summary statistics for the main variables used in the analysis, grouped by agent demographics, agent performance, and customer-agent interaction measures.

Among participating agents, the average age was 32 years, and 18% were male. Figure A1 reports the distribution of agent tenure at the time of the experiment. Most participating agents had started their roles four to five months before treatment.

We construct several performance measures. *Issue identification time* is the elapsed time between the start of a chat and confirmation of the customer issue. Online Appendix C3 describes the chat-stage labeling procedure in detail. *Chat duration* is the total time from the start to the end of a chat. *Customer retrial within three days* is an indicator equal to one if the customer returned with the same issue within three days. On average, issue identification took 35.53 seconds, chat sessions lasted 465.62 seconds, and 38% of customers recontacted the platform for the same issue within three days.

Approximately 15% of chats in the dataset were rated by customers. Figure A2 in the Online Appendix shows that rating response rates for treated and control agents remained stable and similar throughout the experiment. For rated chats, *customer rating* is measured on a one-to-five scale, with higher values indicating better perceived service quality. We also define a *dissatisfaction* indicator equal to one if the customer gave a rating of one or two. The average rating across rated chats during the pretreatment period is 3.46, and 35% of rated chats are classified as dissatisfied.

We also construct interaction measures to capture the customer-agent communication process. *Response time* is the cumulative delay before an agent or customer replies to the other party. We also count the *number of messages* sent by each party. On average, agents sent 11.69 messages and customers sent 11.27 messages per chat. Across all sessions, average response time was 171.29 seconds for agents and 261.31 seconds for customers.

A distinctive feature of this setting is that agents can handle multiple customer chats at the same time and frequently switch across concurrent chats. To measure attention allocation across chats, we use overlapping session timelines to construct three measures. *Concurrency* is the number of simultaneous chats the agent was handling at the start of the focal chat, with a mean of 2.68. A shift-away event occurs when an agent sends a message in the focal chat and the agent's next observed message is sent in another concurrent chat; the interval ends when the agent next sends a message in the focal chat. Based on this procedure, *average cumulative shift-away time* is 181.23 seconds, and *average active time* in the focal chat is 284.65 seconds. Appendix C4 provides the full construction of these measures.

To capture textual characteristics, we compute the total number of words and messages sent by agents and customers in each chat. We also construct three lexical measures based on Read (2000). *Message length* is the average number of Chinese characters per message and captures message elaboration. *Lexical density* is the ratio of content words, such as nouns, verbs, and adjectives, to total words and captures informational richness. *Lexical diversity* is measured using entropy, calculated as $-\sum p\, log_2\, p$, where $p$ represents the frequency of each content word. Higher values indicate greater variation in word usage. These measures are computed separately for agent and customer messages.

We assess randomization by testing balance across 24 key variables. A nonparametric test shows that all p-values for pretreatment variables are above 0.118, indicating no significant differences between treatment and control agents before the experiment. Table A1 in the Online Appendix reports the balance tests. Figure A3 shows parallel pretreatment trends across key outcomes, and Figure A4 compares the pretreatment distributions of customer ratings between treatment and control agents. These results support the validity of the experimental design.

## 4 Empirical Strategy

In this section, we first define dependent variables as service speed and service quality in Section 4.1. We then define independent variables as gen AI access and gen AI usage in Section 4.2. Finally, we introduce our ITT and LATE estimation methods in Sections 4.3 and 4.4, respectively.

### 4.1 Dependent Variables: Service Speed and Service Quality

We first aggregate the chat-level data to the agent-day level. This structure allows us to leverage the panel nature of the data for rigorous difference-in-differences (DID) analyses, which rely on within-agent variation over time. We evaluate the effects of the gen AI assistant using two sets of measures: *service speed* and *service quality*.

***Service speed.*** We measure service speed using two time-based variables: chat duration and issue identification time. Chat duration ($P_c$) represents the total time a customer spends in a chat session. Issue identification time ($P_s$) is the time taken to identify the customer's issue at the beginning of the session. Both measures are commonly used in customer service management to evaluate service efficiency. Shorter durations may indicate greater operational efficiency and lower labor costs, although potentially at the expense of service quality (Goes et al. 2018, Altman et al. 2021).

Figure A5 in the Online Appendix illustrates these speed measures in a hypothetical chat session. The chat starts at 9:00:30 when the agent sends the message "Hello, you are connected to your service representative Henry." The issue is identified at 9:01:30 when the customer responds "Yes, please." The chat ends at 9:09:30 with the message "Nothing else. Thank you!" In this example, issue identification time is one minute, and chat duration is nine minutes.[4]

***Service quality.*** For service quality, we distinguish subjective service quality from objective service quality. Subjective service quality is measured by customer ratings and dissatisfaction rates. Customer rating ($Q_r$) represents the average rating across rated chats for each agent-day. Dissatisfaction rate ($Q_l$) is the proportion of rated chats that received a score below three, that is, one or two on a five-point scale.

Objective service quality is measured by the three-day retrial rate ($Q_d$), defined as the proportion of chats that led to a follow-up contact from the same customer regarding the same issue within three days. This metric is used by Alibaba to assess whether an issue was ultimately resolved. It has also been used in prior research as a proxy for lasting resolution effectiveness (Hu et al. 2022).

### 4.2 Independent Variables: Gen AI Access and Gen AI Usage

In the field experiment, the treatment provided agents with access to the gen AI assistant. Hence, the variable $Treat$ indicates gen AI access at the agent level. However, treated agents have the option to follow or disregard the suggestions of gen AI for each chat session. This feature allows us to record agents' gen AI usage behaviors, e.g., whether and to what extent an agent adheres to gen AI's suggestions. At the chat level, we measure usage using two binary variables: $A_c$ and $A_p$. $A_c = 1$ if an agent directly sends gen AI suggested messages without modifications, i.e., a *complete adherence*, and $A_p = 1$ if an agent copies AI-generated messages into the chat interface and edits them before sending them to the customer, i.e., a *partial adherence*. An agent is considered to have *used* the gen AI assistant in a chat if either $A_c = 1$ or $A_p = 1$, i.e., if the agent completely or partially uses gen AI's suggestions. If both $A_c$ and $A_p$ are zero, the agent disregarded the AI-generated suggestions.

[4] The reason why chat duration can be much longer than issue identification time is because agents need to contact sellers or backend systems to secure approval for the proposed solutions.

Based on these measures, we construct a continuous variable $GenAI_Usage$, defined as the proportion of chats on each agent-day in which an agent either fully or partially adhered to the gen AI assistant (i.e., $A_c = 1$ or $A_p = 1$).[5] Table 2 shows the summary statistics of agent usage of the gen AI assistant. At the agent-day level, the average gen AI usage rate is 0.287, and at the chat level, the average rate is 0.215. The complete adherence rate is 0.219, about three times of the partial adherence rate of 0.070.

To further understand agents' gen AI usage behavior, we plot the distribution of gen AI usage during the treatment period, as shown in Figure A6. The distribution is asymmetric, bimodal, and positively skewed, forming a reverse "J-shape." At the extremes, 17.9% of treated agents did not use the gen AI assistant at all, while 6.2% used it in every assigned chat. Usage rates among the remaining agents are highly heterogeneous.

We also examine the antecedents of gen AI usage across treated agents and chat characteristics. Table A2 reports an OLS analysis using agent characteristics, including gender, age, tenure, pretreatment customer rating, pretreatment concurrency, and pretreatment chat duration. Treated agents with shorter tenure, lower pretreatment concurrency, lower pretreatment customer ratings, and longer pretreatment chat durations tend to use gen AI more. We also examine whether gen AI usage varies across chat contexts. Table A3 shows that usage varies across issue categories and is higher for first-contact chats, but lower when agents handle more concurrent chats. Customer tenure is positively associated with usage, though the magnitude is small.

Finally, we examine how gen AI usage evolved during the treatment period. Figure A7 shows the average daily usage rate among treated agents and reveals a clear upward trend. This increase is consistent with agents becoming more familiar with the tool over time. Figure A8 further shows that complete adherence increased steadily, while partial adherence remained relatively stable throughout the experiment.

Given the incomplete and heterogeneous usage of gen AI in our setting, it is important to distinguish the causal effect of gen AI *access* from the effect of actual gen AI *usage*. We therefore estimate the intention-to-treat (ITT) effect of gen AI access in Section 4.3 and the local average treatment effect (LATE) of gen AI usage in Section 4.4.

### 4.3 Estimating the Intention-to-Treat Effect of Gen AI Access

We first apply the difference-in-differences (DID) method to estimate the ITT effect of gen AI, i.e., the average treatment effect of having access to gen AI, regardless of whether agents used its suggestions. The specification is:

[5] This measure is computed across all chats in a day, regardless of whether they received a customer rating.

$$Y_{it} = a_0 + a_1\, Treat_{it} + \theta_i + \mu_t + \boldsymbol{X}_{i,t-1} + \epsilon_{it}, \quad (1)$$

where $Y \in \{\ln(P_c), \ln(P_s), Q_r, Q_l, Q_d\}$.[6] The subscript $i$ denotes agents, and $t$ denotes calendar days. $\theta_i$ represents agent fixed effects, $\mu_t$ represents day fixed effects, and $\epsilon_{it}$ is the error term. $\boldsymbol{X}_{i,t-1}$ includes a set of lagged time-varying control variables, such as the average daily concurrency and customer tenure experienced by agent $i$ on the previous day. $Treat_{it}$ equals 1 if agent $i$ is in the treatment group and day $t$ falls within the treatment period (i.e., 1/23/2024 – 2/19/2024), and 0 otherwise. The coefficient $a_1$ captures the ITT effect of gen AI access. Robust standard errors are clustered at the agent level for all regressions in this study.

### 4.4 Estimating the Local Average Treatment Effect of Gen AI Usage

The ITT estimates capture the effect of providing access to the gen AI assistant. Because access does not imply use in our setting, we also estimate the effect of actual gen AI usage using an instrumental variable (IV) approach. Specifically, we employ a fuzzy difference-in-differences design, in which the treatment-period gen AI access indicator, $Treat_{it}$, instruments for actual gen AI usage, $GenAI_Usage_{it}$. The panel structure exploits within-agent changes over time, while agent and day fixed effects control for time-invariant agent heterogeneity and common time shocks. This design addresses imperfect compliance and estimates the LATE for agents whose usage increased because they were granted access (de Chaisemartin and D'Haultfoeuille 2018). The empirical specifications are as follows:

*First Stage:*

$$GenAI_Usage_{it} = \alpha_0 + \alpha_1 Treat_{it} + \theta_i + \mu_t + \boldsymbol{X}_{i,t-1} + \nu_{it}, \quad (2)$$

*Second Stage:*

$$Y_{it} = \beta_0 + \beta_1 Gen\widehat{AI_U}sage_{it} + \theta_i + \mu_t + \boldsymbol{X}_{i,t-1} + \epsilon_{it}, \quad (3)$$

where $Y \in \{\ln(P_c), \ln(P_s), Q_r, Q_l, Q_d\}$. The subscripts $i$ and $t$, and the definitions of $Treat_{it}$, $\theta_i$, $\mu_t$, $\boldsymbol{X}_{i,t-1}$ are the same as in Equation (1). $GenAI_Usage_{it}$ indicates the proportion of chats for which agent $i$ used the gen AI assistant on day $t$, including complete and partial adherence.[7] The first stage is strong because access determines whether agents can use the assistant; we report the first-stage estimates with the LATE results. The coefficient of $Gen\widehat{AI_U}sage_{it}$, denoted by $\beta_1$, captures the effect of a one-unit increase in the gen AI usage rate on the outcome variable for compliers, that is, agents whose usage increased as a result of being granted access.

[6] The service speed measures are log-transformed because they have large standard deviations relative to their means.

[7] In our dataset, treatment conditions are randomized at the agent level, while gen AI usage and the outcome measures are observed at the chat level. To leverage the within-unit variation of agents and the panel data structure, we aggregate the independent variable ($GenAI_Usage_{it}$) and all dependent variables ($Y_{it}$) to the agent-day level. Our results are also robust at agent-hour level, as shown in Tables A13 and A14 (Online Appendix).

We use random assignment of gen AI access as an instrument for actual usage to address selection into usage. For example, agents with higher pretreatment customer ratings may be less likely to use the assistant. Because treatment assignment is randomized at the agent level, we instrument for heterogeneous gen AI usage at the agent-day level rather than at the chat level.

For the IV to be valid in the LATE framework, three conditions must hold: *relevance, exclusion, and monotonicity* (Angrist and Imbens 1995, Angrist and Pischke 2009). *Relevance* requires treatment assignment to be associated with gen AI usage. This follows from the experimental design: only treated agents had access to the assistant. *Exclusion* requires the instrument to affect outcomes only through gen AI usage. This condition has two components. The first is independence: the instrument must be unrelated to other factors that affect the outcomes. This is satisfied because $Treat_{it}$ was randomly assigned across agents. The second is no-direct-effect: the IV must not affect outcomes except through gen AI usage. A potential concern is that access to the assistant could affect agents even when they do not use its suggestions. This concern is mitigated by the fact that the assistant appeared in a separate message box and did not alter the original chat workflow. To further assess this concern, we follow Van Kippersluis and Rietveld (2018) and conduct a placebo test among treated agents who never used the assistant, or "never-takers." Table A4 shows no statistically significant effects on key outcomes for this group, supporting the no-direct-effect assumption. Finally, *monotonicity* requires that access to the assistant does not reduce the likelihood of using it, which is naturally satisfied in our setting.

## 5 Impact of Gen AI on Service Performance

In this section, we implement the empirical strategy described in Section 4 to estimate the gen AI assistant's impact on service performance. Section 5.1 presents model-free evidence. Sections 5.2 and 5.3 report the ITT and LATE estimates, respectively. Section 5.4 then examines process-level evidence on how the assistant changes agent-customer interactions.

### 5.1 Model-Free Evidence

We begin with model-free evidence by comparing the pretreatment and treatment-period distributions of the main dependent variables for treated agents, reported in Online Appendix Figure A9. The distributions are consistent with improvements in service speed and subjective service quality after the gen AI assistant is deployed: issue identification time, chat duration, and dissatisfaction shift leftward, while customer ratings shift rightward. By contrast, the three-day retrial rate shows little visible change. The corresponding means show similar patterns. Average issue identification time decreases from 41.18 seconds to 38.61 seconds, and average chat duration falls from 534.22 seconds to 507.58 seconds. Customer ratings improve from 3.16 to 3.27, while dissatisfaction rates decline from 0.43 to 0.40. The retrial rate remains stable at 0.39. These descriptive patterns motivate the regression analyses below, where we formally estimate the causal effects of access to, and use of, the gen AI assistant.

### 5.2 Gen AI's ITT Effect

Table 3 presents the estimated ITT effect of access to the gen AI assistant. For service speed, Columns (1) and (2) show that access to the gen AI assistant reduced issue identification time and chat duration by 8.2% and 1.1%, respectively, corresponding to absolute reductions of 3.2 seconds and 5.7 seconds.

For subjective service quality, Columns (3) and (4) show that access to the assistant decreased the dissatisfaction rate by 0.012 (a 3.4% improvement relative to the pretreatment mean of 0.35) and raised average customer ratings by 0.042 points on a five-point scale (a 1.2% improvement over the pretreatment mean of 3.46).[8] Column (5) shows a near-zero and statistically insignificant effect on three-day retrial rate. All estimates except the retrial rate estimate are statistically significant. Overall, providing access to the gen AI assistant improved service speed and subjective service quality, while leaving objective resolution largely unchanged.

### 5.3 Gen AI's LATE Effect

The ITT estimates capture the causal effect of providing access to the gen AI assistant. Because access does not imply use in our setting, we next estimate the local average treatment effect (LATE) of actual gen AI usage, reported in Tables 4A and 4B. In both tables, Column (1) shows a statistically significant first stage, using treatment assignment as an instrument for gen AI usage.

For service speed, the second-stage estimates in Table 4A Columns (2) and (3) show that an increase in the gen AI usage rate from 0 to 1 decreases issue identification time by 32.3% and chat duration by 4.2%. For subjective service quality, Table 4B Columns (2) and (3) show that an increase in usage from 0 to 1 decreases the dissatisfaction rate by 0.054, a 15.4% improvement relative to the pretreatment mean of 0.35, and raises average customer ratings by 0.184 points on a five-point scale, a 5.3% improvement relative to the pretreatment mean of 3.46. Column (4) reports a near-zero and statistically insignificant effect on three-day retrial rate. Overall, the LATE estimates are roughly four times as large as the corresponding ITT estimates.

In summary, the ITT and LATE results show that gen AI improves service speed and subjective service quality, but not objective service quality. This pattern is consistent with the institutional setting. Alibaba's SOPs define what agents are authorized to do, and these rules did not change during the experiment. The gen AI assistant may help agents diagnose issues more quickly and communicate more effectively, thereby improving customers' service experience. However, it does not expand the set of available solutions or give agents additional authority to resolve issues outside existing platform rules. Thus, gen AI can improve service interaction without necessarily changing whether the underlying customer issue is ultimately resolved.

---

[8] We estimated the gen AI assistant's impact on customers' likelihood to rate the chat, and the effect is close to zero. See Table A5 for details.

### 5.4 Mechanisms for Gen AI's Impact on Service Performance

By design, the gen AI assistant can directly improve service performance by helping agents diagnose customer issues and formulate solution proposals. These functions shorten the early part of the chat, reducing issue identification time and overall chat duration, as shown above. We next examine whether these gains reflect only this automation benefit or whether gen AI also changes agent-customer interaction.

To examine the interaction process, we divide each chat into three stages: *issue identification and solution proposal*, *resolution*, and *confirmation* (see Online Appendix C3 for details). Although the assistant is deployed primarily in the first stage, its effects on the interaction process may extend into later stages if agents use the initial AI support to organize the conversation more effectively. We therefore examine four sets of process measures: *agent response times*, *agent typing time*, *agent-customer message ratios*, and *customer communication measures*. Detailed variable definitions and examples are provided in Online Appendix C.

We begin with <u>*agent response time*</u>. Table 5 Column (1) shows that a one-unit increase (i.e., from 0 to 1) in the gen AI usage rate reduces agent response time for the entire chat by 5.7%. This reduction is strongest in the issue identification and solution proposal stage, where the assistant is directly involved. Table 5 Column (2) shows that agent response time in this stage decreases by 10.0% ($p<0.01$). Importantly, the response time reduction also appears beyond the first stage when gen AI is deployed. Table 5 Column (4) shows that agent response time decreases by 8.0% in the third stage ($p<0.05$). These results suggest that gen AI helps agents move the conversation forward faster, both during the initial diagnosis and in later parts of the chat. By contrast, the estimated effects on customer response time are negative but statistically insignificant across the full chat and each stage, indicating that the main acceleration effect operates on the agent side.

We next examine whether faster responses reflect lower <u>*agent effort*</u>. Table 6 Column (1) shows that gen AI usage has no significant effect on agent typing time. This result provides little evidence that agents become faster simply by withdrawing effort or relying passively on AI-generated text. Instead, the communication balance shifts toward greater agent input.[9] Table 6 Column (2) shows that gen AI significantly increases the agent-customer message ratio over the full chat. This increase is concentrated in the issue identification and solution proposal stage, where the assistant directly supports diagnosis and solution formulation, and remains significant in the confirmation stage. These patterns suggest that gen AI makes agents more proactive in service interaction: they respond faster while taking a larger role in the communication process.

[9] To measure communication balance, we construct the agent-customer message ratio, defined as the number of agent messages divided by the number of customer messages.

Finally, we examine whether this greater agent proactivity reduces *customers' contribution to the interaction*. Table 7 shows that customers interacting with gen AI-assisted agents provide less demanding forms of input. Gen AI usage significantly increases the number and share of short affirmative replies, such as brief confirmations or acknowledgments, and significantly reduces customer-uploaded pictures. Because picture uploads typically require customers to provide visual evidence or diagnostic information, the decline in pictures suggests that customers need to do less work to complete the conversation. Table 7 Columns (4)–(6) further show that customer messages become shorter and less lexically complex, consistent with reduced customer contribution.

As supplementary evidence on communication style, Online Appendix Table A6 reports textual characteristics by chat stage. We interpret these results cautiously because AI can directly affect message wording. In the issue identification and solution proposal stage, where gen AI suggestions are offered, agent messages become longer and lexically richer, while customer messages become shorter and less lexically complex. In later stages, agent-side textual effects are no longer statistically significant, while customer messages remain shorter and less lexically diverse.

In summary, these results suggest that the performance gains induced by gen AI come from more than automation. Assisted by gen AI, age*nts become more proactive in servic*e: they respond faster and take a larger role in the service interaction, both across the full chat and in later stages where the assistant is no longer deployed. This shift in agent-customer interaction reduces the contribution required from customers, who respond with more short affirmations, upload fewer pictures, and send shorter, less lexically complex messages. These patterns suggest that gen AI reshapes the service interaction rather than merely automating diagnosis and solution formulation.

## 6 Heterogeneous Effects of Gen AI on Service Performance

So far, we have presented causal evidence showing that gen AI improves average service performance. These average effects, however, may mask substantial heterogeneity across agents. A central concern in AI deployment is that the same tool may affect workers differently depending on their baseline skill. To examine this issue, we estimate heterogeneous effects by pretreatment performance of agents, using average pretreatment customer ratings as a proxy for skill. Agents are classified into quintiles (Q1–Q5), with Q1 representing the lowest-performing agents and Q5 the highest.[10]

We first report the heterogeneous ITT effects of gen AI access. As shown in Table A7, gen AI usage declines with pretreatment performance, from 39.3% for Q1 agents to 18.3% for Q5 agents, suggesting that lower-performing agents incorporated the gen AI tool more readily, whereas top performers used it more conservatively. We therefore also estimate the heterogeneous LATE effects of actual gen AI usage. The ITT and LATE results are reported in Tables 8 and 9, respectively.[11] The LATE estimates capture the effect of a one-unit increase in the agent-day usage rate among agents whose usage was induced by access. Multiplying these estimates by the actual usage rate of each quintile gives the implied effect under observed usage, which we compare with the ITT estimates.

In Section 6.1, we examine the heterogeneous effects of gen AI by pretreatment performance. In Section 6.2, we investigate mechanisms that may explain these heterogeneous effects.

### 6.1 Heterogeneous Effects Across Pretreatment Performance Quintiles

#### 6.1.1 Heterogeneous Effects of Gen AI on Service Speed

We begin with service speed, measured by issue identification time and chat duration. Table 8 Columns (1)–(2) report the ITT estimates. Access to the gen AI assistant reduced issue identification time across all quintiles. Specifically, issue identification time decreases by 6.6%, 11.0%, 11.5%, 7.4%, and 4.0% for Q1 through Q5 agents, respectively. However, the effects on overall chat duration are more limited. Gen AI access significantly reduces chat duration only for Q1 and Q3 agents, by 5.9% and 2.3%, respectively.

We next turn to the LATE estimates, reported in Table 9, which capture the effect of actual gen AI usage induced by access. These estimates are directionally consistent with the ITT results but are larger in magnitude, as expected under incomplete usage. For *issue identification time*, gen AI improves agent performance across all quintiles. As shown in Column (1), a one-unit increase in the agent-day usage rate reduces issue identification time by 16.8% ($p<0.05$) for Q1 agents, 36.2% ($p<0.05$) for Q2 agents, 46.2% ($p<0.01$) for Q3 agents, 37.8% ($p<0.01$) for Q4 agents, and 18.3% ($p<0.05$) for Q5 agents. This pattern suggests that the largest gains arise among mid-tier performers, with smaller gains for both the lowest- and highest-performing agents.

For *chat duration*, the effects are concentrated among lower-performing agents. Table 9 Column (2) shows that a one-unit increase in the agent-day usage rate shortens chat duration by 14.8% for Q1 agents (p<0.01), 5.3% for Q2 agents (p>0.1), and 9.0% for Q3 agents (p<0.01), whereas Q4 and Q5 agents show no statistically significant changes.

---

[10] For each agent, we calculate their average customer rating during the pretreatment period. Then, we classify agents into performance groups. Although our classification is based on customer ratings, the results are robust to using other performance measures such as service speed.

[11] For the heterogeneous LATE analysis, we follow Angrist and Pischke (2009) to use $Treat_{it}$ and $Treat_{it} \times Agent_Type_i$ as the IVs for endogenous variables (i.e., $GenAI_Usage_{it}$ and $GenAI_Usage_{it} \times Agent_Type_i$) in the second stage regression.

Scaling the LATE estimates by the actual usage rates in Table A7 yields implied issue identification time reductions of 6.6%, 9.2%, 10.1%, 6.8%, and 3.3% for Q1 through Q5 agents, respectively. These magnitudes are close to the corresponding ITT estimates in Table 8: 6.6%, 11.0%, 11.5%, 7.4%, and 4.0%. For chat duration, the scaled LATE estimates for Q1–Q3 agents are reductions of 5.8%, 1.3%, and 2.0%, respectively, again close to the ITT estimates of 5.9%, 1.6%, and 2.3%. The ITT and LATE estimates provide consistent evidence that gen AI improves issue identification speed for all agents, while reductions in overall chat duration are concentrated among lower-performing agents.

### 6.1.2 Heterogeneous Effects of Gen AI on Service Quality

We next examine the heterogeneous effects on service quality, distinguishing subjective service quality from objective resolution. Subjective service quality is measured by customer ratings and dissatisfaction rates, whereas objective service quality is measured by the three-day customer retrial rate.

Columns (3)–(5) of Table 8 report the ITT estimates. For subjective service quality, access to the gen AI assistant generates the largest gains for Q1 agents, reducing dissatisfaction by 0.215 (p<0.01) and increasing customer ratings by 0.812 points (p<0.01). These benefits decline with higher pretreatment performance but remain positive and statistically significant for Q2 agents (-0.051 for dissatisfaction and 0.185 for ratings, both p<0.01) and Q3 agents (-0.021 for dissatisfaction and 0.076 for ratings, both p<0.01). The effects become statistically insignificant for Q4 agents and reverse for Q5 agents, whose dissatisfaction rate increases by 0.070 (p<0.01) and whose ratings fall by 0.283 points (p<0.01). For objective resolution, the three-day retrial estimates are small and statistically insignificant for Q1–Q4 agents, but Q5 agents experience a 0.009 increase in retrials (p<0.05). These results suggest that access to gen AI improves subjective service quality for lower-performing agents, while quality deteriorates for top performers.

Model-free comparisons in Table A8 are consistent with the ITT evidence: average customer ratings increase from the pretreatment to posttreatment period for Q1–Q4 agents but decline for Q5 agents. This evidence suggests that gen AI narrows performance gaps while preserving the observed ranking of service quality across quintiles.

The LATE estimates reveal a sharper pattern than the ITT estimates. For *subjective service quality*, the effects exhibit a clear monotonic pattern across performance levels (Table 9 Columns (3)–(4)). A one-unit increase in the agent-day usage rate benefits Q1 agents the most: their dissatisfaction rate decreases by 0.567 (p<0.01), and their customer ratings increase by 2.144 points (p<0.01). These effects decline steadily with higher baseline performance among Q2-Q5 agents. The quality effects become statistically insignificant for Q4 agents. Surprisingly, gen AI usage hurt Q5 agents' quality, as shown by an increase in dissatisfaction rate by 0.374 ($p<0.01$) and a decrease in customer ratings by 1.514 points ($p<0.01$).

For *objective service quality*, gen AI usage has no statistically significant effects on Q1–Q4 agents (Table 9 Column (5)). The only exception is Q5, where the customer retrial rate increases by 0.041 ($p<0.05$), again indicating a quality decline among these top performers.

The LATE estimates again closely track the ITT patterns. For both dissatisfaction rates and customer ratings, the LATE estimates show the same monotonic shift in effects across pretreatment performance quintiles: large gains for Q1 agents, smaller but positive gains for Q2 and Q3 agents, little effect for Q4 agents, and negative effects for Q5 agents. For objective service quality, both approaches show small and statistically insignificant effects for Q1–Q4 agents but a positive retrial effect for Q5 agents. Thus, the quality results show a robust pattern: gen AI improves subjective service quality for lower-performing agents, has little effect for Q4 agents, and worsens both subjective and objective service quality for Q5 agents.

### 6.2 Mechanisms Behind the Heterogeneous Effects of Gen AI

The heterogeneous analysis in Section 6.1 identifies two main performance patterns following the introduction of the gen AI assistant. First, the *service speed gap narrows*: low performers show significant reduction in chat duration, while high performers show limited or no change. All groups, however, benefit from faster issue identification. Second, the *service quality gap also narrows*: low performers achieve notable gains in service quality, whereas top performers experience a significant decline. We next examine the mechanisms behind these heterogeneous effects.

#### 6.2.1 Mechanism for the Service Speed Gap Reduction

The mechanism for the reduction in service speed gap is relatively direct. The gen AI assistant is designed to support the issue identification and solution proposal stage by generating diagnostic and solution suggestions. This function reduces the time agents need to interpret customer requests and formulate initial replies, which explains why issue identification time decreases across all quintiles. The effects on overall chat duration, however, are concentrated among lower-performing agents. These agents likely faced greater baseline frictions in initial diagnosis, solution formulation, and chat management, leaving more room for the assistant to shorten the overall interaction. By contrast, high performers already handled chats efficiently before treatment, automating the early stage produced little additional reduction in total chat duration.

#### 6.2.2 Mechanism for the Service Quality Gap Reduction

The mechanism for service quality is less direct. Unlike service speed, which can improve when the assistant shortens issue identification and solution formulation, service quality depends on how agents integrate gen AI's suggestions into their multi-chat workflow. Results in Section 6.1.2 show that actual gen AI usage improves subjective service quality for lower-performing agents but worsens both subjective and objective service quality for top performers. This pattern raises a central question: *why would using the same gen AI assistant benefit lower performers but hurt top performers?* In the following, we will propose a primary mechanism and provide suggestive evidence to support it. We will also explore two alternative mechanisms that are not supported by our observed evidence.

**Primary Mechanism: Workflow disruption among top performers.** *We conjecture that gen AI disrupts the established workflow of top performers by increasing shift-away time from the focal chat and delaying responses to customers.*

We examine two links in the workflow disruption mechanism. The first is whether gen AI usage changes how Q5 agents allocate attention across concurrent chats. The second link is whether this change in attention allocation lowers service quality in the focal chat. Because shift-away behavior is observational rather than randomly assigned, we interpret it as suggestive evidence rather than as a formally identified causal mediator of quality decline. We therefore examine whether shift-away behavior, response delays, and the timing of customer retrials form a pattern consistent with weakened service continuity.

We construct shift-away measures from observed message transitions across concurrent chats. A shift-away event begins when an agent's next observed message after a focal-chat message is sent in another concurrent chat and ends when the agent returns to the focal chat. Table 10 shows that Q5 agents are the only group whose overall shift-away time increases, by 23.8% ($p<0.1$). The increase occurs after the stage in which the gen AI assistant is directly deployed: shift-away time rises by 21.9% in the resolution stage ($p<0.1$) and by 27.3% in the confirmation stage ($p<0.05$). By contrast, shift-away time decreases for Q1, Q3, and Q4 agents, while the estimate for Q2 agents is negative but statistically insignificant.

To understand this chat-switching behavior better, we analyze the destination of these shifts, as shown in Table A12. Among Q5 agents, 50.1% of shift-away episodes lead to the issue identification and solution proposal stage of the destination chat, compared with 27.1% leading to the destination chat's resolution stage and 22.8% to its confirmation stage. Thus, the recorded shift-away episodes involve observable work in other customer chats, especially the diagnosis or initiation of the destination chat.

Prior research explains why chat switching may weaken service continuity and reduce service quality. Multitasking in online customer service can lengthen in-service delays and reduce problem resolution, both of which lower customer satisfaction (Goes et al. 2018). Interruptions can also require workers to reconstruct task-specific knowledge when they return to an interrupted case (Froehle and White 2014), and recent evidence documents performance costs from sustained multitasking in live-chat service (Batt and Gallino 2025). These studies provide a basis for conjecturing the implications of increased shift-away time.

The response time and retrial results follow the same pattern. Table 11 Columns (1)–(4) show that gen AI reduces agent response time for Q1–Q3 agents but increases it for Q5 agents. For Q5 agents, overall response time rises by 8.6% ($p<0.05$), with larger increases in the resolution stage (15.5%, $p<0.1$) and confirmation stage (17.4%, $p<0.05$). These are the same stages in which Q5 shift-away time increases. Table 11 Columns (5)–(6) further show that customer retrials within ten minutes increase by 0.035 ($p<0.01$), whereas retrials occurring between ten minutes and three days increase by only 0.007 and are not statistically significant.

In sum, the LATE estimates show that gen AI usage changes Q5 agents' observed allocation of attention across concurrent chats. The destination, response time, and customer retrial results are consistent with this attention reallocation that weakens the service continuity of the focal chat.

**Alternative Conjecture 1:** *Top performers receive weaker AI diagnostics than low performers.*

An alternative conjecture is that Q5 agents receive more difficult cases or cases for which AI performs less well than low performers. The matching algorithm uses historical performance, so agents in different performance quintiles may operate in different task environments. Persistent differences across quintiles do not by themselves bias the Q5 treatment estimate: random assignment compares treated and control Q5 agents operating under the same matching regime. Nevertheless, we conduct two analyses to test the validity of this alternative conjecture.

Specifically, we assess observable task assignment and measured AI diagnostic consistency in two ways. First, Table A9 shows that the distribution of broad issue categories remains similar between the pretreatment and treatment periods within each performance quintile. Second, we construct an issue identification match indicator that records whether the issue inferred from the AI suggestion matches the category recorded by the agent at the end of the chat. Table A10 shows no evidence that this consistency is lower for Q5 chats: the Q5-Q1 difference is -0.006 and statistically insignificant in the full sample, and it remains insignificant within the reported chat subsamples. Table A11 also shows that Q5 agents' lower adherence persists within these subsamples.

Taken together, these results do not support the conjecture that Q5 agents perform worse because they receive harder chats or because the AI provides less reliable diagnoses for their chats. We find no systematic differences in broad issue categories and AI-agent diagnostic match between Q5 agents and low performers.

**Alternative Conjecture 2:** *Overreliance on low-quality AI suggestions.*

A second possible reason for top performers' quality decline is that they rely too often or too indiscriminately on low-quality AI suggestions. The observed usage patterns do not support this conjecture. As shown in Table A7, gen AI usage declines from 39.3% for Q1 agents to 18.3% for Q5 agents. Table A11 further shows that Q5 adherence is 0.163 lower than Q1 adherence in the full sample ($p<0.01$), and this difference persists within the reported chat subsamples. Survey responses in Online Appendix E also indicate that high performers evaluate the gen AI assistant more skeptically.

Agents in all performance groups adopt gen AI suggestions selectively. Table A20 in Online Appendix F shows that adoption is higher when a suggestion's inferred issue matches the issue category recorded at the end of the chat. For Q5 agents, the adoption rate is 19.7% for consistent suggestions and 15.5% for inconsistent ones, a gap of 4.2 percentage points; the corresponding rates for Q1 agents are 40.4% and 37.0%. Because these gaps are similar across quintiles, the evidence does not show that Q5 agents are better at detecting potential AI errors. It does show that they do not accept suggestions indiscriminately, at least no less selectively than agents in other quintiles.

Greater skepticism toward AI suggestions could instead lead Q5 agents to spend more time verifying suggestions before responding. The decision time analyses shown in Tables A21 and A22 of Online Appendix F do not support this explanation. Q5 agents take 6.41 seconds on average to adopt or reject an AI suggestion, compared with 7.71 seconds for Q1 agents, and the adjusted Q5-Q1 difference is small and statistically insignificant. Table 11 likewise shows no significant increase in Q5 response time during the initial issue identification and solution proposal stage, where the gen AI assistant is deployed. The adverse process effects appear later, during the resolution and confirmation stages.

These findings suggest that neither overreliance on gen AI nor extended verification within the focal chat is the main explanation for top performers' quality decline. The process evidence is more consistent with gen AI disrupting their existing workflow routines of serving concurrent chats.

## 7 Conclusion

This study examines how a generative AI assistant changes service performance and the organization of service work in a large-scale randomized field experiment on the Alibaba e-commerce platform. The assistant provides issue diagnoses and solution proposals as text suggestions that agents can adopt, modify, or disregard. This setting allows us to estimate both the effect of gen AI access and the effect of actual use in a high-volume, concurrent-chat service environment. We find that gen AI improves service speed and subjective service quality, as reflected in shorter issue identification time, shorter chat duration, higher customer ratings, and lower dissatisfaction. However, it does not improve objective service quality, measured by customer retrials.

The process evidence shows that performance gains do not arise solely from automation in the stage when the assistant is deployed. Although the tool supports issue identification and solution proposal during the opening stage of the chat, its effects extend into later stages of the service interaction. Gen AI helps agents respond faster, shifts communication toward greater agent input, and reduces the input required from customers. At the same time, the absence of a change in agent typing time provides little evidence that faster service results from reduced agent communication activity. Taken together, these patterns suggest that gen AI changes the structure of agent-customer interaction. In this setting, its value lies not only in producing text or diagnoses, but also in reshaping the service process and customer experience.

The effects are also uneven across workers. Lower-performing agents benefit the most, especially in speed and perceived service quality, suggesting that gen AI can standardize service delivery and narrow performance gaps. Top-performing agents, however, experience declines in both subjective service quality and objective resolution. The evidence is most consistent with a workflow-disruption interpretation. In a concurrent-chat environment, gen AI use among top performers increases their shift-away time, slows their responses during resolution and confirmation, and raises customers' immediate retrials. These patterns do not suggest that top performers overly rely on gen AI. If anything, they use it more conservatively and adhere to it less often. Instead, the results suggest that introducing gen AI into an already efficient human workflow can alter attention allocation across concurrent chats and weaken service continuity.

These findings contribute to research on AI and service operations by clarifying where gen AI creates value, for whom it creates or destroys value, and how it changes service process. Managerially, our findings suggest that firms should view gen AI deployment as a process design problem rather than simply a technology adoption decision. Gen AI can improve service speed and perceived service quality, but these gains may not translate into better objective resolution. Managers should therefore evaluate deployment using both perception-based and resolution-based outcomes. The heterogeneous effects also caution against uniform rollout: while gen AI may help lower-performing agents standardize service delivery and narrow performance gaps, firms should preserve discretion for top-performing agents and monitor whether AI use disrupts attention allocation and service continuity.

Several limitations suggest directions for future research. First, because agents choose when and how to use the assistant, our study cannot identify the optimal level of AI usage. Second, our mechanism evidence is based on process measures and should be interpreted as consistent with workflow disruption, rather than as definitive proof of a single causal channel. Third, future research could examine how gen AI affects long-run learning or deskilling, how AI tools should be designed for expert workers, and how firms should allocate control rights as these systems become more autonomous. More broadly, our findings suggest that the productivity value of gen AI depends on how well organizations align the tool with existing processes and performance goals.

## References

**Acemoglu, D., Anderson, G. W., Beede, D. N., Buffington, C., Childress, E. E., Dinlersoz, E., ... & Zolas, N. (2022a).** Automation and the workforce: A firm-level view from the 2019 Annual Business Survey. Technical Report w30659, *National Bureau of Economic Research*.

**Acemoglu, D., Autor, D., Hazell, J., & Restrepo, P. (2022b).** Artificial intelligence and jobs: Evidence from online vacancies. *Journal of Labor Economics*, 40(S1), S293-S340.

**Agrawal, A., Gans, J. S., & Goldfarb, A. (2023).** Do we want less automation? AI may provide a path to decrease inequality. *Science*, 381 (6654), 155-158.

**Agrawal, N., Moehring, A., Rajpurkar, P., & Salz, T. (2023).** Combining human expertise with artificial intelligence: Experimental evidence from radiology. Technical Report w31422, *National Bureau of Economic Research*.

**Altman, D., Yom-Tov, G. B., Olivares, M., Ashtar, S., & Rafaeli, A. (2021).** Do customer emotions affect agent speed? An empirical study of emotional load in online customer contact centers. *Manufacturing & Service Operations Management*, 23(4), 854-875.

**Angrist, J. D., & Imbens, G. (1995).** Identification and estimation of local average treatment effects. *National Bureau of Economic Research*.

**Angrist, J. D., & Pischke, J. S. (2009).** Mostly harmless econometrics: An empiricist's companion. *Princeton University Press.*

**Autor, D. H. (2024).** Applying AI to rebuild middle class jobs. Technical Report w32140, *National Bureau of Economic Research*.

**Bai, B., Dai, H., Zhang, D. J., Zhang, F., & Hu, H. (2022).** The impacts of algorithmic work assignment on fairness perceptions and productivity: Evidence from field experiments. *Manufacturing & Service Operations Management*, 24(6), 3060-3078.

**Balakrishnan, M., Ferreira, K. J., & Tong, J. (2026).** Human-Algorithm collaboration with private information: Naïve advice weighting behavior and mitigation. *Management Science*, 72(1), 265-284.

**Bastani, H., Bastani, O., & Sinchaisri, W. P. (2026).** Improving human sequential decision making with reinforcement learning. *Management Science*, 72(1), 733-755.

**Batt, R., & Gallino, S. (2025).** Multitasking over time: The time-dependent effects of multitasking. Available at *SSRN*: 4823942.

**Boyaci, T., Canyakmaz, C., & De Véricourt, F. (2024).** Human and machine: The impact of machine input on decision making under cognitive limitations. *Management Science*, 70(2), 1258-1275.

**Brynjolfsson, E., Li, D., & Raymond, L. (2025).** Generative AI at work. *The Quarterly Journal of Economics*, 140(2), 889-942.

**Capraro, V., Lentsch, A., Acemoglu, D., Akgun, S., Akhmedova, A., Bilancini, E., ... & Viale, R. (2024).** The impact of generative artificial intelligence on socioeconomic inequalities and policy making. *PNAS nexus*, 3(6).

**Caro, F., & de Tejada Cuenca, A. S. (2023).** Believing in analytics: Managers' adherence to price recommendations from a DSS. *Manufacturing & Service Operations Management*, 25(2), 524-542.

**Castelo, N., Bos, M. W., & Lehmann, D. R. (2019).** Task-dependent algorithm aversion. *Journal of marketing research*, 56(5), 809-825.

**Cui, R., Li, M., & Zhang, S. (2022).** AI and procurement. *Manufacturing & Service Operations Management*, 24(2), 691-706.

**De Chaisemartin, C., & D'Haultfoeuille, X. (2018).** Fuzzy differences-in-differences. *The Review of Economic Studies*, 85(2), 999-1028.

**Dell'Acqua, F. (2022).** Falling asleep at the wheel: Human/AI collaboration in a field experiment on HR recruiters. *Working Paper*, Laboratory for Innovation Science, Harvard Business School.

**Dell'Acqua, F., McFowland III, E., Mollick, E. R., Lifshitz-Assaf, H., Kellogg, K., Rajendran, S., ... & Lakhani, K. R. (2026).** Navigating the jagged technological frontier: Field experimental evidence of the effects of artificial intelligence on knowledge worker productivity and quality. *Organization Science*, 37(2), 403-423.

**Dietvorst, B. J., Simmons, J. P., & Massey, C. (2018).** Overcoming algorithm aversion: People will use imperfect algorithms if they can (even slightly) modify them. *Management Science*, 64(3), 1155-1170.

**Froehle, C. M., & White, D. L. (2014).** Interruption and forgetting in knowledge-intensive service environments. *Production and Operations Management*, 23(4), 704-722.

**Goes, P. B., Ilk, N., Lin, M., & Zhao, J. L. (2018).** When more is less: Field evidence on unintended consequences of multitasking. *Management Science*, 64(7), 3033-3054.

**Grönroos, C. (1984).** A service quality model and its marketing implications. *European Journal of Marketing*, 18(4), 36-44.

**Hosseini Maasoum, S. M., & Lichtinger, G. (2025).** Generative AI as seniority-biased technological change: Evidence from US résumé and job posting data. Available at *SSRN*: 5425555.

**Hou, T., Li, M., Tan, Y., & Zhao, H. (2024).** Physician adoption of AI assistant. *Manufacturing & Service Operations Management*, 26(5), 1639-1655.

**Hu, K., Allon, G., & Bassamboo, A. (2022).** Understanding customer retrials in call centers: Preferences for service quality and service speed. *Manufacturing & Service Operations Management*, 24(2), 1002-1020.

**Huang, M.-H., & Rust, R. T. (2018).** Artificial intelligence in service. *Journal of Service Research*, 21(2), 155-172.

**Karau, S. J., & Williams, K. D. (1993).** Social loafing: A meta-analytic review and theoretical integration. *Journal of personality and social psychology*, 65(4), 681.

**Lebovitz, S., Lifshitz-Assaf, H., & Levina, N. (2022).** To engage or not to engage with AI for critical judgments: How professionals deal with opacity when using AI for medical diagnosis. *Organization science*, 33(1), 126-148.

**Lehmann, C. A., Haubitz, C. B., Fügener, A., & Thonemann, U. W. (2022).** The risk of algorithm transparency: How algorithm complexity drives the effects on the use of advice. *Production and Operations Management*, 31(9), 3419-3434.

**Luo, X., Tong, S., Fang, Z., & Qu, Z. (2019).** Frontiers: Machines vs. humans: The impact of artificial intelligence chatbot disclosure on customer purchases. *Marketing Science*, 38(6), 937-947.

**McElheran, K., Li, J. F., Brynjolfsson, E., Kroff, Z., Dinlersoz, E., Foster, L., & Zolas, N. (2024).** AI adoption in America: Who, what, and where. *Journal of Economics & Management Strategy*, 33(2), 375-415.

**Noy, S., & Zhang, W. (2023).** Experimental evidence on the productivity effects of generative artificial intelligence. *Science*, 381(6654), 187-192.

**Parasuraman, A., Zeithaml, V. A., & Berry, L. L. (1985).** A conceptual model of service quality and its implications for future research. *Journal of Marketing*, 49(4), 41-50.

**Peng, S., Kalliamvakou, E., Cihon, P., & Demirer, M. (2023).** The impact of AI on developer productivity: Evidence from GitHub Copilot. *arXiv* preprint arXiv:2302.06590.

**Read, J. A. (2000).** Assessing vocabulary. *Cambridge University Press*.

**Rubinstein, J. S., Meyer, D. E., & Evans, J. E. (2001).** Executive control of cognitive processes in task switching. *Journal of Experimental Psychology: Human Perception and Performance*, 27(4), 763.

**Schanke, S., Burtch, G., & Ray, G. (2021).** Estimating the impact of "humanizing" customer service chatbots. *Information Systems Research*, 32(3), 736-751.

**Snyder, C., Keppler, S., & Leider, S. (2026).** Algorithm reliance: Fast and slow. *Management Science,* 72(1), 368-385.

**Sun, J., Zhang, D. J., Hu, H., & Van Mieghem, J. A. (2022).** Predicting human discretion to adjust algorithmic prescription: A large-scale field experiment in warehouse operations. *Management Science*, 68(2), 846-865.

**Van Donselaar, K. H., Gaur, V., Van Woensel, T., Broekmeulen, R. A., & Fransoo, J. C. (2010).** Ordering behavior in retail stores and implications for automated replenishment. *Management Science*, 56(5), 766-784.

**Van Kippersluis, H., & Rietveld, C. A. (2018).** Beyond plausibly exogenous. *The Econometrics Journal*, 21(3), 316-331.

**Waardenburg, L., Huysman, M., & Sergeeva, A. V. (2022).** In the land of the blind, the one-eyed man is king: Knowledge brokerage in the age of learning algorithms. *Organization science*, 33(1), 59-82.

**Xu, Y., Dai, H., & Yan, W. (2026).** Identity disclosure and anthropomorphism in voice chatbot design: A field experiment. *Management Science*, 72(1), 223-241.

**Zhang, S., & Narayandas, D. (2026).** Engaging customers with AI in online chats: Evidence from a randomized field experiment. *Management Science*, 72(1), 73-95.

**Zolas, N., Kroff, Z., Brynjolfsson, E., McElheran, K., Beede, D. N., Buffington, C., Goldschlag, N., Foster, L., & Dinlersoz, E. (2021).** Advanced technologies adoption and use by US firms: Evidence from the annual business survey. Technical Report w28290, *National Bureau of Economic Research*.

## Table 1. Summary Statistics

| Variables | Definition | Obs. | Mean | SD |
|---|---|---|---|---|
| **Agent Demographics** | | | | |
| treat | 1 if agent was exposed to the gen AI assistant, and 0 otherwise | 5,940 | 0.49 | 0.50 |
| gender | 1 if agent is male, and 0 otherwise | 5,940 | 0.18 | 0.38 |
| tenure | total time since an agent began their role (days) | 5,940 | 123.84 | 53.94 |
| age | agent age | 5,940 | 31.75 | 6.10 |
| **Agent Performance** | | | | |
| ***Service Speed*** | | | | |
| identification time | time taken to identify the customer's issue (seconds) | 2,507,730 | 35.53 | 70.66 |
| chat duration | total duration of a chat (seconds) | 2,564,606 | 465.62 | 447.15 |
| ***Service Quality*** | | | | |
| customer rating | customer rating of the chat, from 1 to 5 with higher number being better | 388,793 | 3.49 | 1.80 |
| if dissatisfied | 1 if the chat receives a rating of 1 or 2 | 388,793 | 0.34 | 0.47 |
| customer retrial | 1 if the customer returns with the same issue within 3 days | 2,564,606 | 0.38 | 0.49 |
| **Agent-Customer Interaction Measures** | | | | |
| ***Agent Measures*** | | | | |
| concurrency | number of chats an agent serves simultaneously when starting a focal chat | 2,564,265 | 2.68 | 1.07 |
| agent response time | total delay before agent replies to customer messages (seconds) | 2,564,600 | 171.29 | 178.04 |
| agent active time | total time an agent spends on the focal chat (seconds) | 2,563,039 | 284.65 | 291.10 |
| agent shift-away time | total time an agent is away from the focal chat (seconds) | 2,563,039 | 181.23 | 246.62 |
| agent typing time | total time an agent spends typing messages (seconds) | 2,532,419 | 105.02 | 111.46 |
| number of agent messages | total number of messages sent by the agent during a chat | 2,564,606 | 11.69 | 8.26 |
| average agent message length | average characters per agent message during a chat | 2,558,467 | 15.70 | 7.28 |
| agent lexical density | ratio of content words to total words in agent messages during a chat | 2,558,463 | 0.50 | 0.09 |
| agent lexical diversity | entropy of content words in agent messages during a chat | 2,558,467 | 4.42 | 1.26 |
| ***Customer Measures*** | | | | |
| customer tenure | total time since a customer joined platform (years) | 2,564,606 | 7.48 | 3.77 |
| customer response time | total delay before customer replies to agent messages (seconds) | 2,564,600 | 261.31 | 303.77 |
| number of customer messages | total number of messages sent by the customer during a chat | 2,564,606 | 11.27 | 10.98 |
| average customer message length | average characters per customer message during a chat | 2,550,289 | 8.85 | 11.69 |
| number of customer pictures | total number of pictures sent by the customer during a chat | 2,559,053 | 0.89 | 1.74 |
| customer lexical density | ratio of content words to total words in customer messages during a chat | 2,549,030 | 0.56 | 0.14 |
| customer lexical diversity | entropy of content words in customer messages during a chat | 2,550,289 | 3.75 | 1.24 |

## Table 2. Agent Usage of Gen AI: Summary Statistics

| Variables | Obs. | Mean | SD |
|---|---|---|---|
| *Agent-Day Level Statistics for Treated Agents* | | | |
| GenAI Usage Rate | 29,006 | 0.287 | 0.363 |
| Complete Adherence Rate | 29,006 | 0.219 | 0.339 |
| Partial Adherence Rate | 29,006 | 0.070 | 0.198 |
| *Chat Level Statistics for Treated Agents* | | | |
| GenAI Usage Rate | 543,651 | 0.215 | 0.411 |
| Complete Adherence Rate | 543,651 | 0.171 | 0.377 |
| Partial Adherence Rate | 543,651 | 0.044 | 0.206 |

***Notes.*** In our setting, "complete adherence" occurs when an agent adopts unmodified genAI outputs in a chat; "partial adherence" occurs when an agent uses part of genAI outputs in a chat by editing the text. GenAI usage includes both complete and partial adherence.

## Table 3. Gen AI's ITT Effects

| | Service Speed | | Service Quality | | |
|---|---|---|---|---|---|
| | ln(identification time) | ln(chat duration) | dissatisfaction rate | customer rating | retrial rate |
| | (1) | (2) | (3) | (4) | (5) |
| Treat | -0.082*** | -0.011** | -0.012*** | 0.042** | 0.000 |
| | (0.012) | (0.005) | (0.005) | (0.018) | (0.002) |
| Agent Fixed Effects | Y | Y | Y | Y | Y |
| Day Fixed Effects | Y | Y | Y | Y | Y |
| Time-Variant Controls | Y | Y | Y | Y | Y |
| Observations | 125,821 | 126,365 | 97,354 | 97,354 | 126,365 |
| R-Squared | 0.335 | 0.236 | 0.159 | 0.164 | 0.084 |

***Notes.*** Observations in Column (1) are less than in Column (2) because some customers or agents may quit the chat before or immediately after agent assignment. In Column (3), dissatisfied customers rated the chat 1-2 stars. *p<0.1; **p<0.05; ***p<0.01. Robust standard errors are clustered by agent.

## Table 4A. Gen AI's LATE Effects on Service Speed

| | First Stage | Second Stage | |
|---|---|---|---|
| | GenAI_Usage | ln(identification time) | ln(chat duration) |
| | (1) | (2) | (3) |
| Treat | 0.254*** | | |
| | (0.007) | | |
| GenAI_Usage | | -0.323*** | -0.042** |
| | | (0.046) | (0.020) |
| Agent Fixed Effects | Y | Y | Y |
| Day Fixed Effects | Y | Y | Y |
| Time-Variant Controls | Y | Y | Y |
| Observations | 126,365 | 125,821 | 126,365 |

***Notes.*** Observations in Columns (2) are less than in Columns (1) and (3) because some customers or agents may quit before or immediately after agent assignment. *p<0.1; **p<0.05; ***p<0.01. Robust standard errors are clustered at the agent level.

## Table 4B. Gen AI's LATE Effects on Service Quality

| | First Stage | Second Stage | | |
|---|---|---|---|---|
| | GenAI_Usage | dissatisfaction rate | customer rating | retrial rate |
| | (1) | (2) | (3) | (4) |
| Treat | 0.228*** | | | |
| | (0.007) | | | |
| GenAI_Usage | | -0.054*** | 0.184** | 0.000 |
| | | (0.021) | (0.079) | (0.009) |
| Agent Fixed Effects | Y | Y | Y | Y |
| Day Fixed Effects | Y | Y | Y | Y |
| Time-Variant Controls | Y | Y | Y | Y |
| Observations | 97,354 | 97,354 | 97,354 | 126,365 |

*Notes.* Customer ratings are on a five-star scale and dissatisfied customers' ratings are below three stars. *p<0.1; **p<0.05; ***p<0.01. Robust standard errors are clustered at the agent level.

## Table 5. (Mechanism) Gen AI's LATE Effects on Response Time by Chat Stage

| | ln(agent response time) | | | |
|---|---|---|---|---|
| | *entire chat* | *chat stage 1* issue identification & solution proposal | *chat stage 2* resolution | *chat stage 3* confirmation |
| | (1) | (2) | (3) | (4) |
| GenAI_Usage | -0.057*** | -0.100*** | -0.031 | -0.080** |
| | (0.021) | (0.029) | (0.040) | (0.036) |
| Agent Fixed Effects | Y | Y | Y | Y |
| Day Fixed Effects | Y | Y | Y | Y |
| Time-Variant Controls | Y | Y | Y | Y |
| Observations | 126,357 | 126,347 | 122,419 | 122,444 |
| | **ln(customer response time)** | | | |
| | *entire chat* | *chat stage 1* issue identification & solution proposal | *chat stage 2* resolution | *chat stage 3* confirmation |
| | (5) | (6) | (7) | (8) |
| GenAI_Usage | -0.018 | -0.013 | -0.015 | -0.037 |
| | (0.024) | (0.042) | (0.040) | (0.035) |
| Agent Fixed Effects | Y | Y | Y | Y |
| Day Fixed Effects | Y | Y | Y | Y |
| Time-Variant Controls | Y | Y | Y | Y |
| Observations | 126,299 | 125,886 | 122,418 | 122,954 |

*Notes.* *p<0.1; **p<0.05; ***p<0.01. Robust standard errors are clustered at the agent level.

## Table 6. (Mechanism) Gen AI's LATE Effects on Agent Typing Time and Agent-Customer Message Ratio by Chat Stage

| | agent typing time | agent-customer message ratio | | | |
|---|---|---|---|---|---|
| | *entire chat* | *entire chat* | *chat stage 1* issue identification & solution proposal | *chat stage 2* resolution | *chat stage 3* confirmation |
| | (1) | (2) | (3) | (4) | (5) |
| GenAI_Usage | 0.005 | 0.139*** | 0.321*** | 0.019 | 0.100*** |
| | (0.006) | (0.013) | (0.015) | (0.016) | (0.033) |
| Agent Fixed Effects | Y | Y | Y | Y | Y |
| Day Fixed Effects | Y | Y | Y | Y | Y |
| Time-Variant Controls | Y | Y | Y | Y | Y |
| Observations | 126,361 | 126,348 | 126,274 | 122,023 | 122,726 |

*Notes.* *p<0.1; **p<0.05; ***p<0.01. Robust standard errors are clustered at the agent level.

## Table 7. (Mechanism) Gen AI's LATE Effects on Customer Input

| | customer textual and communication characteristics | | | | | |
|---|---|---|---|---|---|---|
| | # of short affirmative replies | share of short affirmative replies | # of pictures | message length | lexical density | lexical diversity |
| | (1) | (2) | (3) | (4) | (5) | (6) |
| GenAI_Usage | 0.201*** | 0.041*** | -0.102*** | -0.833*** | -0.017*** | -0.108*** |
| | (0.027) | (0.003) | (0.039) | (0.191) | (0.003) | (0.024) |
| Agent Fixed Effects | Y | Y | Y | Y | Y | Y |
| Day Fixed Effects | Y | Y | Y | Y | Y | Y |
| Time-Variant Controls | Y | Y | Y | Y | Y | Y |
| Observations | 126,365 | 126,365 | 126,352 | 126,348 | 126,347 | 126,348 |

***Notes.*** *p<0.1; **p<0.05; ***p<0.01. Robust standard errors are clustered at the agent level.

## Table 8. Gen AI's Heterogeneous ITT Effects: Quintile Partition of Agent Pretreatment Performance

| | **Service Speed** | | **Service Quality** | | |
|---|---|---|---|---|---|
| | ln(identification time) | ln(chat duration) | dissatisfaction rate | customer rating | retrial rate |
| | (1) | (2) | (3) | (4) | (5) |
| Treat | -0.066** | -0.059*** | -0.215*** | 0.812*** | 0.000 |
| | (0.033) | (0.015) | (0.016) | (0.059) | (0.007) |
| Treat × Q2 Agents | -0.044 | 0.043** | 0.163*** | -0.626*** | -0.001 |
| | (0.039) | (0.018) | (0.018) | (0.068) | (0.008) |
| Treat × Q3 Agents | -0.050 | 0.037** | 0.194*** | -0.735*** | -0.002 |
| | (0.036) | (0.015) | (0.016) | (0.062) | (0.007) |
| Treat × Q4 Agents | -0.008 | 0.062*** | 0.221*** | -0.835*** | -0.003 |
| | (0.035) | (0.015) | (0.016) | (0.060) | (0.007) |
| Treat × Q5 Agents | 0.025 | 0.072*** | 0.285*** | -1.095*** | 0.009 |
| | (0.037) | (0.016) | (0.017) | (0.065) | (0.008) |
| Impact on Q1 Agents | -0.066** | -0.059*** | -0.215*** | 0.812*** | 0.000 |
| Impact on Q2 Agents | -0.110*** | -0.016 | -0.051*** | 0.185*** | -0.001 |
| Impact on Q3 Agents | -0.115*** | -0.023*** | -0.021*** | 0.076*** | -0.002 |
| Impact on Q4 Agents | -0.074*** | 0.003 | 0.007 | -0.024 | -0.003 |
| Impact on Q5 Agents | -0.040** | 0.013 | 0.070*** | -0.283*** | 0.009** |
| Agent Fixed Effects | Y | Y | Y | Y | Y |
| Day Fixed Effects | Y | Y | Y | Y | Y |
| Time-Variant Controls | Y | Y | Y | Y | Y |
| Observations | 122,055 | 122,554 | 95,561 | 95,561 | 122,554 |

***Notes.*** *p<0.1; **p<0.05; ***p<0.01. Robust standard errors are in parentheses.

## Table 9. Gen AI's Heterogeneous LATE Effects: Quintile Partition of Agent Pretreatment Performance

| | **Service Speed** | | **Service Quality** | | |
|---|---|---|---|---|---|
| | ln(identification time) | ln(chat duration) | dissatisfaction rate | customer rating | retrial rate |
| | (1) | (2) | (3) | (4) | (5) |
| GenAI_Usage | -0.168** | -0.148*** | -0.567*** | 2.144*** | -0.000 |
| | (0.082) | (0.037) | (0.052) | (0.199) | (0.018) |
| GenAI_Usage × Q2 Agents | -0.194* | 0.096* | 0.390*** | -1.507*** | -0.003 |
| | (0.107) | (0.050) | (0.062) | (0.236) | (0.023) |
| GenAI_Usage × Q3 Agents | -0.294*** | 0.058 | 0.479*** | -1.823*** | -0.007 |
| | (0.107) | (0.044) | (0.058) | (0.221) | (0.021) |
| GenAI_Usage × Q4 Agents | -0.210* | 0.162*** | 0.603*** | -2.270*** | -0.017 |
| | (0.110) | (0.048) | (0.058) | (0.223) | (0.022) |
| GenAI_Usage × Q5 Agents | -0.015 | 0.207*** | 0.941*** | -3.657*** | 0.041* |
| | (0.117) | (0.053) | (0.070) | (0.272) | (0.024) |
| Impact on Q1 Agents | -0.168** | -0.148*** | -0.567*** | 2.144*** | -0.000 |
| Impact on Q2 Agents | -0.362** | -0.053 | -0.177*** | 0.636*** | -0.003 |
| Impact on Q3 Agents | -0.462*** | -0.090*** | -0.088*** | 0.321*** | -0.007 |
| Impact on Q4 Agents | -0.378*** | 0.013 | 0.037 | -0.126 | -0.017 |
| Impact on Q5 Agents | -0.183** | 0.059 | 0.374*** | -1.514*** | 0.041** |
| Agent Fixed Effects | Y | Y | Y | Y | Y |
| Day Fixed Effects | Y | Y | Y | Y | Y |
| Time-Variant Controls | Y | Y | Y | Y | Y |
| Observations | 122,055 | 122,554 | 95,561 | 95,561 | 122,554 |

***Notes.*** *p<0.1; **p<0.05; ***p<0.01. Robust standard errors are in parentheses.

### Table 10. (Mechanism) Gen AI's Heterogeneous LATE Effects on Agent Shift-Away Time

| | ln(agent shift-away time) | | | |
|---|---|---|---|---|
| | | *chat stage 1* | *chat stage 2* | *chat stage 3* |
| | *entire chat* | issue identification & solution proposal | resolution | confirmation |
| | (1) | (2) | (3) | (4) |
| GenAI_Usage | -0.298** | -0.387*** | -0.124 | -0.313** |
| | (0.125) | (0.112) | (0.119) | (0.132) |
| GenAI_Usage × Q2 Agents | 0.181 | -0.053 | -0.002 | 0.199 |
| | (0.159) | (0.159) | (0.152) | (0.163) |
| GenAI_Usage × Q3 Agents | 0.019 | 0.035 | 0.028 | 0.198 |
| | (0.147) | (0.135) | (0.135) | (0.156) |
| GenAI_Usage × Q4 Agents | 0.043 | 0.345** | 0.216 | 0.134 |
| | (0.154) | (0.146) | (0.142) | (0.158) |
| GenAI_Usage × Q5 Agents | 0.536*** | 0.553*** | 0.343** | 0.586*** |
| | (0.165) | (0.162) | (0.162) | (0.175) |
| Impact on Q1 Agents | -0.298** | -0.387*** | -0.124 | -0.313** |
| Impact on Q2 Agents | -0.117 | -0.440*** | -0.126 | -0.114 |
| Impact on Q3 Agents | -0.189** | -0.351*** | -0.096 | -0.115 |
| Impact on Q4 Agents | -0.255** | -0.042 | 0.092 | -0.179* |
| Impact on Q5 Agents | 0.238* | 0.166 | 0.219* | 0.273** |
| Agent Fixed Effects | Y | Y | Y | Y |
| Day Fixed Effects | Y | Y | Y | Y |
| Time-Variant Controls | Y | Y | Y | Y |
| Observations | 113,230 | 103,771 | 99,909 | 108,354 |

***Notes.*** *p<0.1; **p<0.05; ***p<0.01. Robust standard errors are in parentheses.

### Table 11. (Mechanism) Gen AI's Heterogeneous LATE Effects on Agent Response Time and Customer Retrials

| | ln(agent response time) | | | | Customer | |
|---|---|---|---|---|---|---|
| | | *chat stage 1* | *chat stage 2* | *chat stage 3* | | |
| | *entire chat* | issue identification & solution proposal | resolution | confirmation | retrial rate (within 10 mins) | retrial rate (10 mins to 3 days) |
| | (1) | (2) | (3) | (4) | (5) | (6) |
| GenAI_Usage | -0.130*** | -0.084* | -0.057 | -0.080 | -0.002 | 0.002 |
| | (0.041) | (0.045) | (0.080) | (0.072) | (0.013) | (0.016) |
| GenAI_Usage × Q2 Agents | 0.067 | -0.059 | 0.022 | -0.088 | 0.011 | -0.015 |
| | (0.053) | (0.058) | (0.105) | (0.091) | (0.016) | (0.021) |
| GenAI_Usage × Q3 Agents | -0.007 | -0.088 | -0.019 | -0.052 | -0.003 | -0.004 |
| | (0.049) | (0.061) | (0.092) | (0.085) | (0.015) | (0.019) |
| GenAI_Usage × Q4 Agents | 0.113** | 0.069 | -0.037 | -0.046 | 0.003 | -0.019 |
| | (0.052) | (0.069) | (0.100) | (0.090) | (0.015) | (0.020) |
| GenAI_Usage × Q5 Agents | 0.217*** | 0.038 | 0.212* | 0.254*** | 0.037** | 0.004 |
| | (0.057) | (0.069) | (0.115) | (0.098) | (0.016) | (0.022) |
| Impact on Q1 Agents | -0.130*** | -0.084* | -0.057 | -0.080 | -0.002 | 0.002 |
| Impact on Q2 Agents | -0.063* | -0.143*** | -0.035 | -0.169*** | 0.009 | -0.012 |
| Impact on Q3 Agents | -0.137*** | -0.172*** | -0.076 | -0.132** | -0.005 | -0.002 |
| Impact on Q4 Agents | -0.017 | -0.015 | -0.095 | -0.125** | 0.000 | -0.017 |
| Impact on Q5 Agents | 0.086** | -0.046 | 0.155* | 0.174** | 0.035*** | 0.007 |
| Agent Fixed Effects | Y | Y | Y | Y | Y | Y |
| Day Fixed Effects | Y | Y | Y | Y | Y | Y |
| Time-Variant Controls | Y | Y | Y | Y | Y | Y |
| Observations | 122,549 | 122,539 | 119,043 | 119,025 | 122,554 | 122,554 |

***Notes.*** *p<0.1; **p<0.05; ***p<0.01. Robust standard errors are in parentheses.

# Generative AI in Action: Field Experimental Evidence from Alibaba's Customer Service Operations

## Online Appendix

This *Online Appendix* provides supplementary evidence and measurement details for the main paper.

- *Appendix A* reports supplementary figures on sample characteristics, randomization checks, usage patterns, and descriptive evidence.
- *Appendix B* reports supplementary tables on balance, gen AI usage, robustness checks, heterogeneous patterns, and additional analyses.
- *Appendix C* defines the agent-customer interaction measures used in the mechanism analyses.
- *Appendix D* explains the construction of the issue identification matching measure used to assess whether the gen AI assistant's diagnostic suggestions align with agents' final issue classifications.
- *Appendix E* reports evidence from an anonymous online survey of treated agents.

### Appendix A. Figures

**Figure A1. Distribution of Agent Tenure at the Time of Experiment**

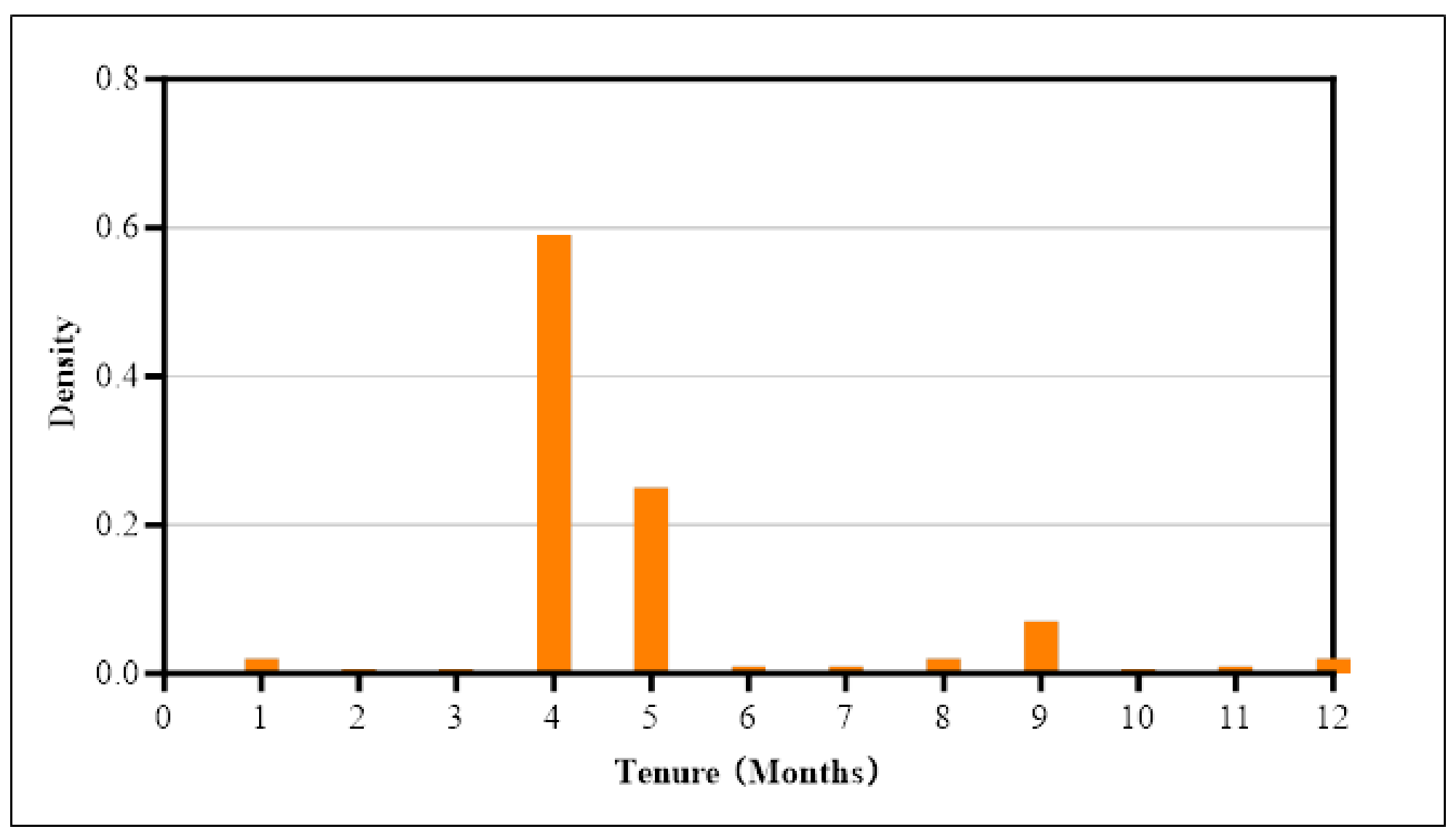

**Figure A2. Customer Rating Response Rates for Treated and Control Agents Over Time**

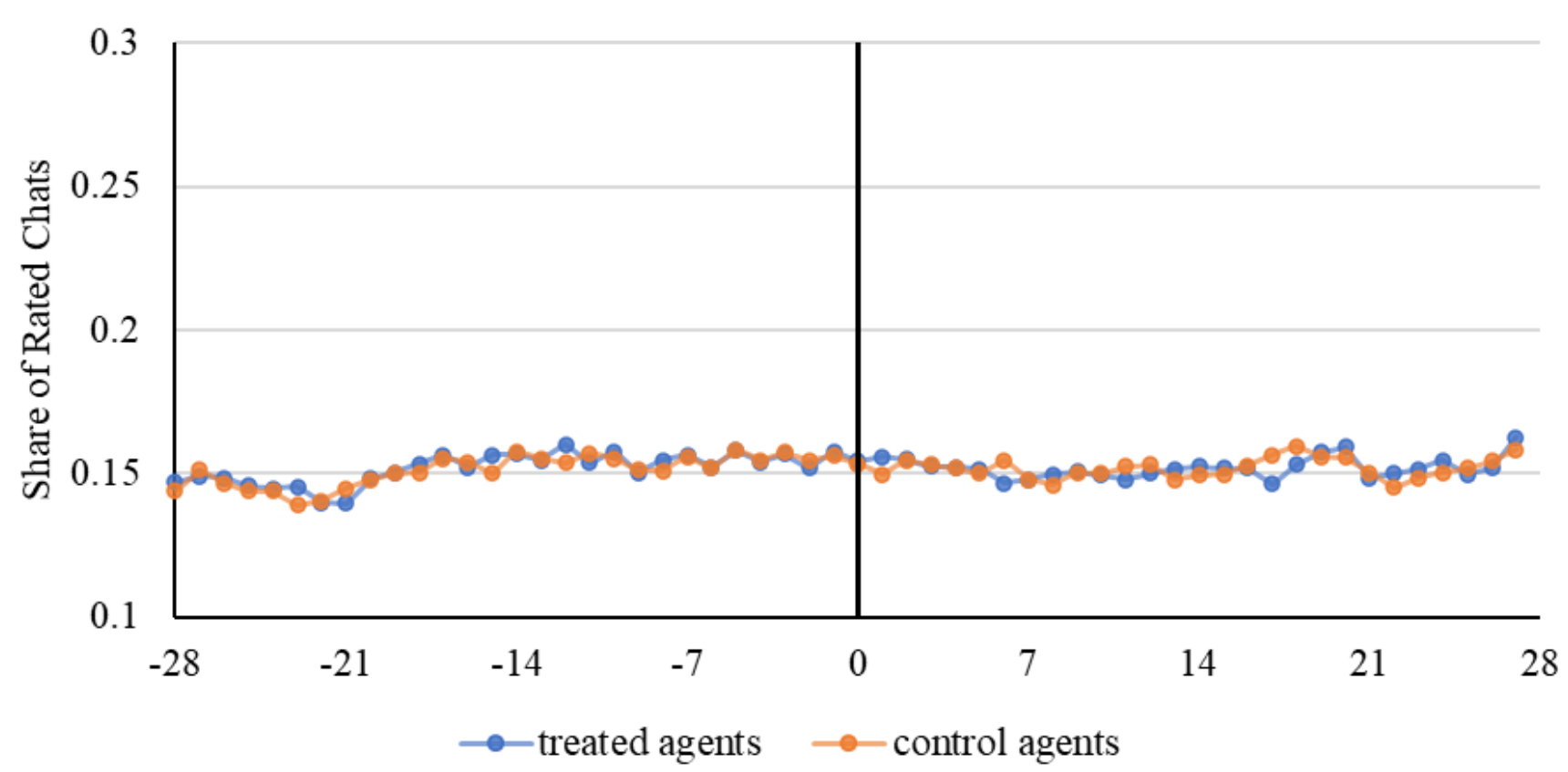

**Figure A3. Parallel Pretreatment Trends**

Figure A3-A. Visual Evidence of Parallel Pretreatment Trends

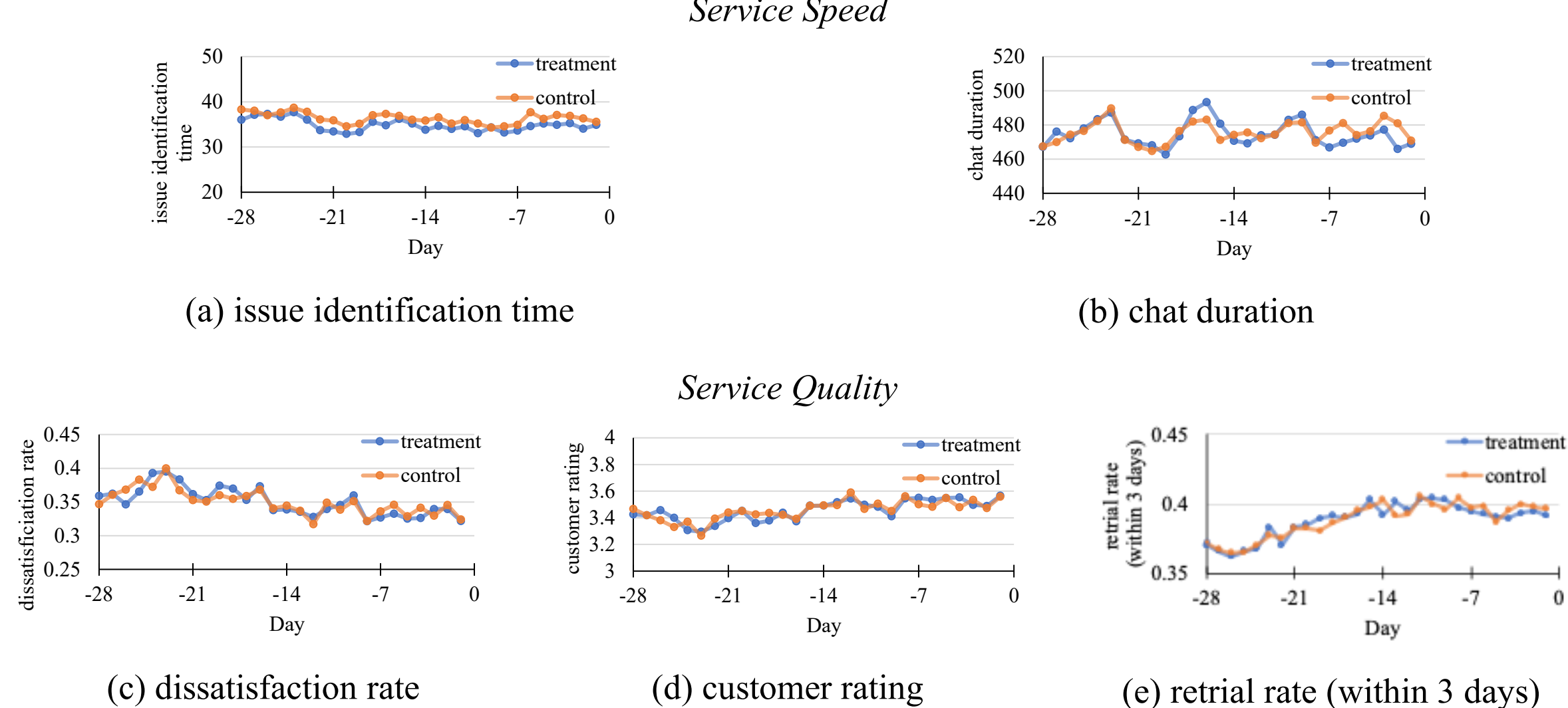

(a) issue identification time

(b) chat duration

(c) dissatisfaction rate

(d) customer rating

(e) retrial rate (within 3 days)

***Notes.*** These figures present the daily average performance of agents in the treatment and control groups during the pretreatment period. The x-axis denotes days relative to the treatment start date (Day 0), and the y-axis indicates average agent performance at the chat level. As a time-sensitive metric, it displays periodic fluctuations.

Figure A3-B. Statistical Tests of Parallel Pretreatment Trends

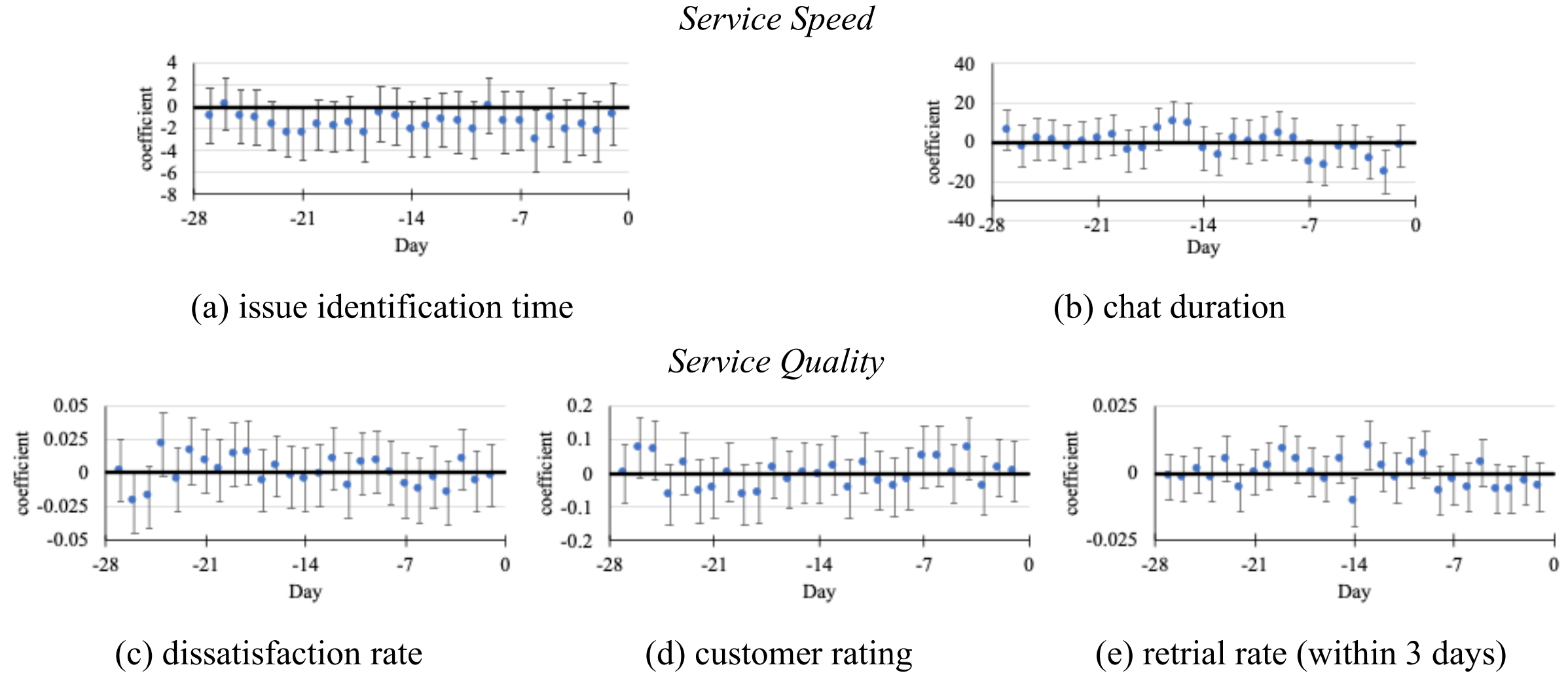

***Notes.*** These plots show the coefficients and 95% confidence intervals for the parallel pre-trend tests.

**Figure A4. Pretreatment Distribution of Customer Ratings for Treated and Control Agents**

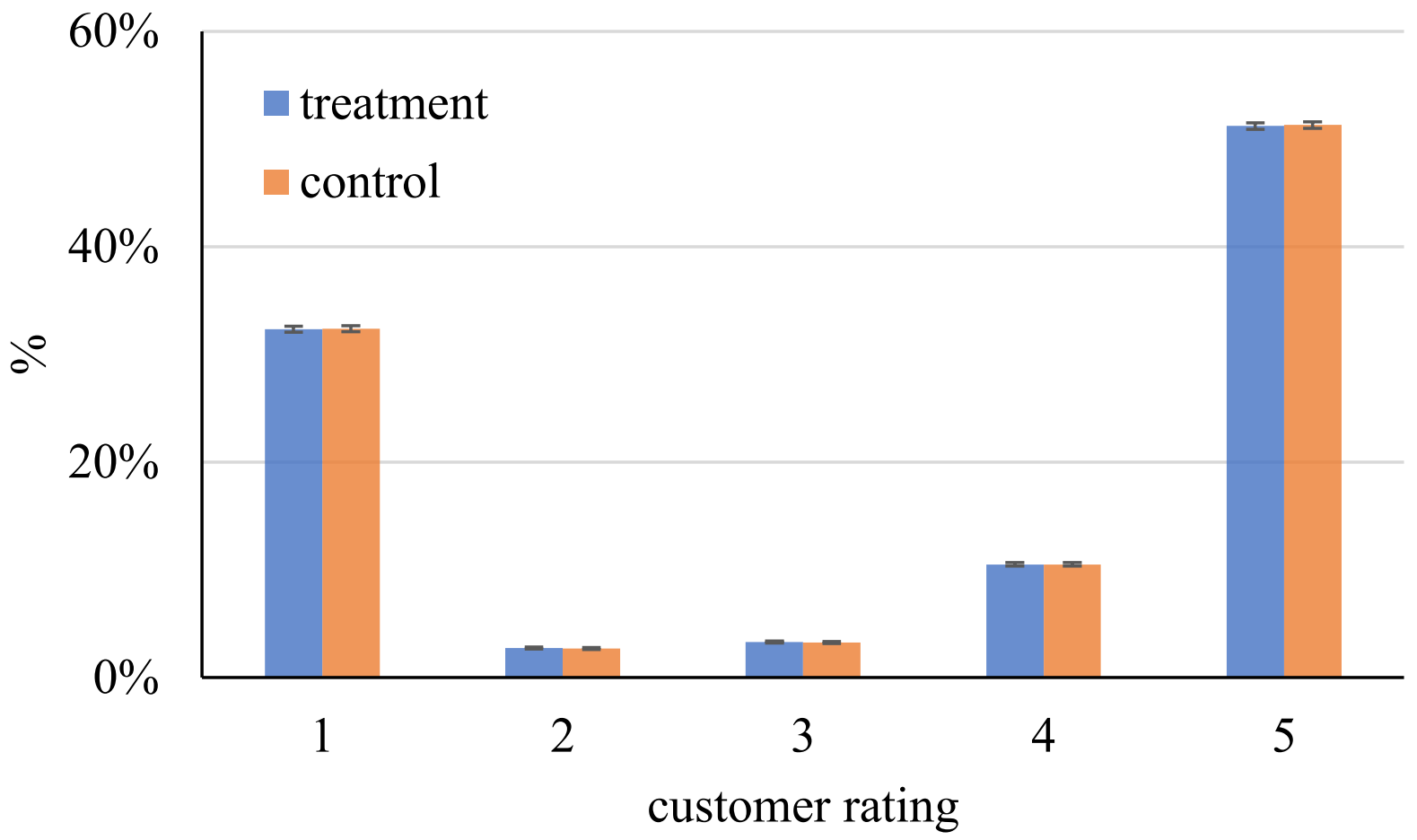

## Figure A5. Process Measures for a Hypothetical Chat Session

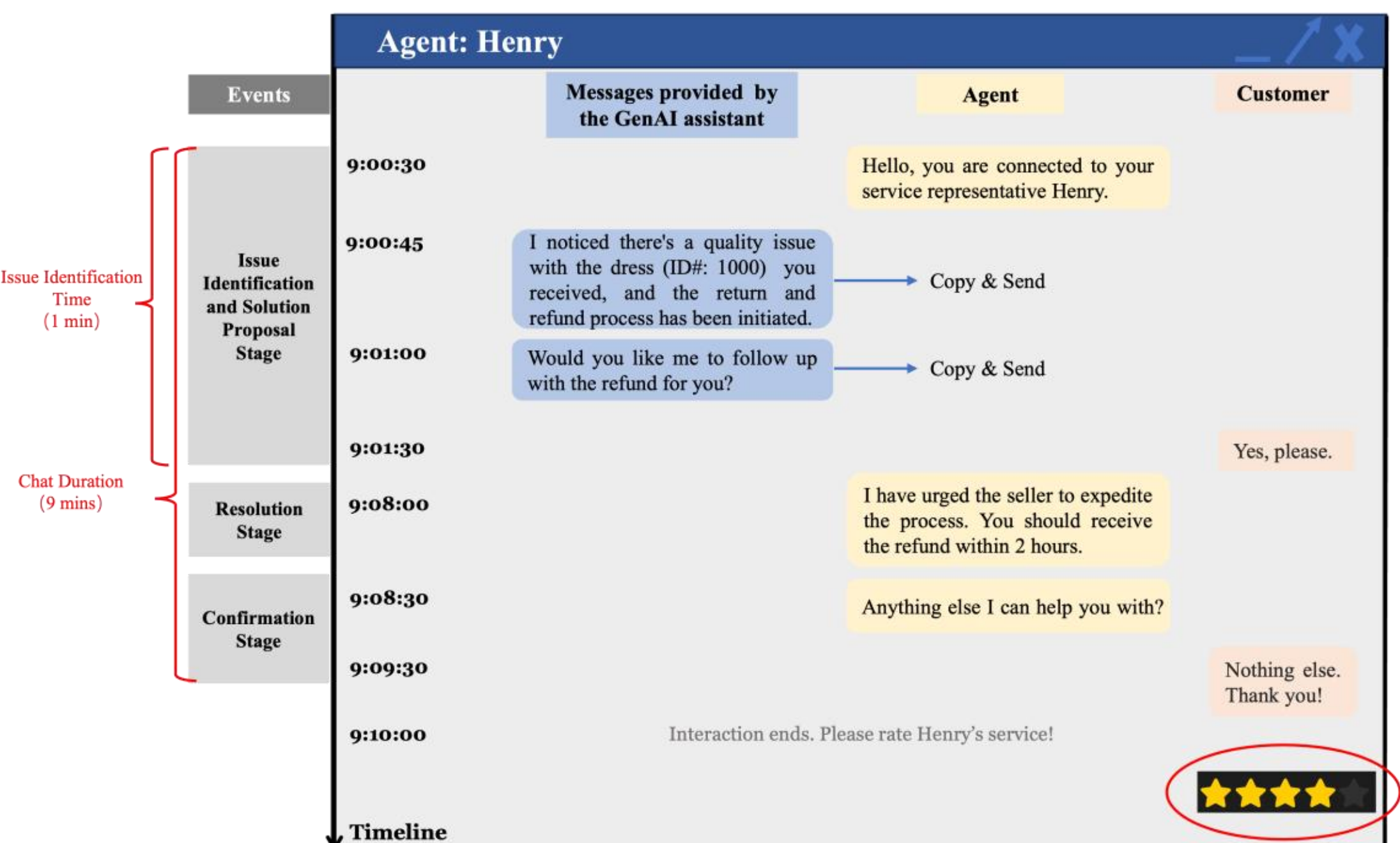

## Figure A6. Distribution of Treated Agents' Gen AI Usage

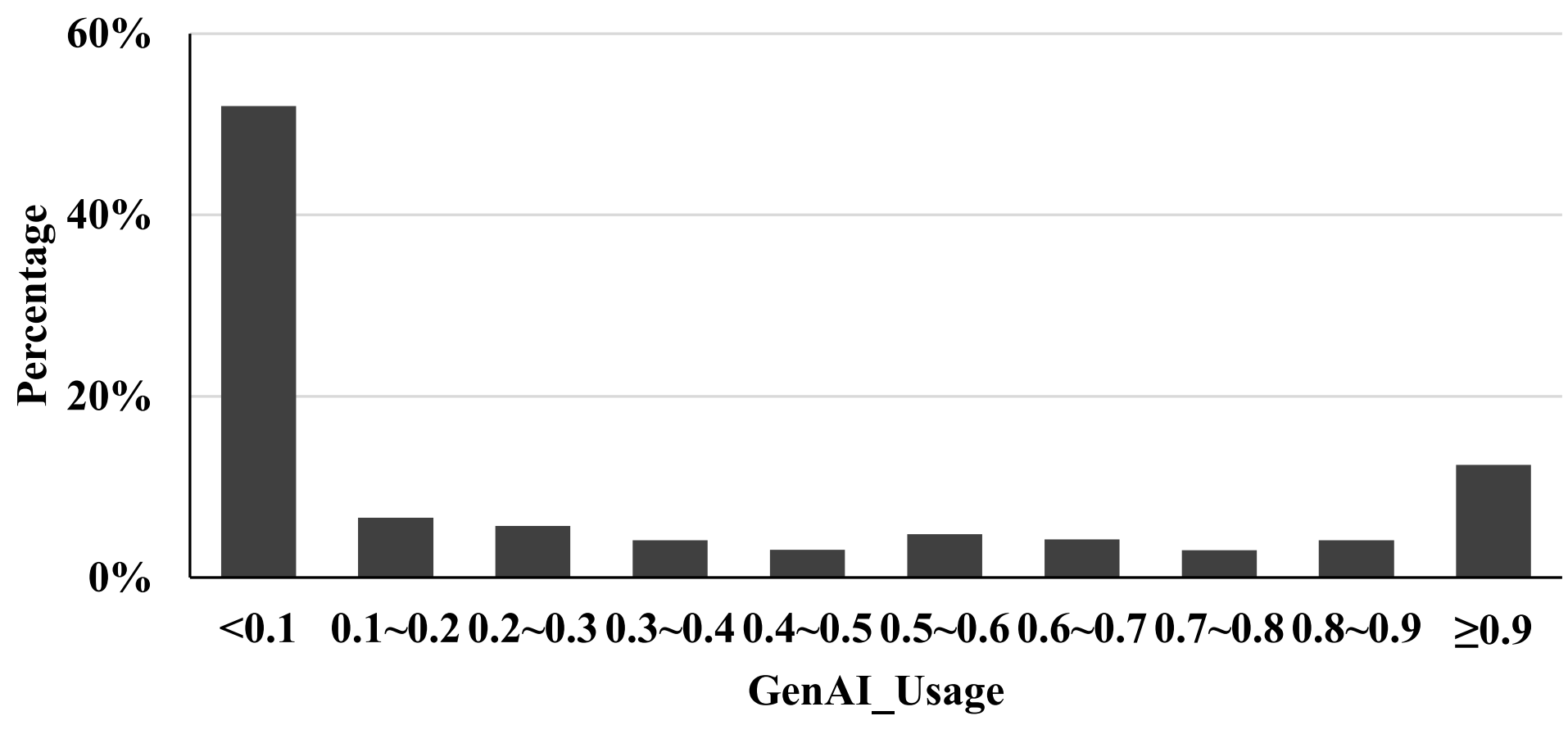

***Notes.*** Gen AI usage includes both complete adherence (i.e., use gen AI recommendations with no modification) and partial adherence (i.e., use modified gen AI recommendations). Usage rates are reported at the agent-day level.

**Figure A7. Average Gen AI Usage Over Time**

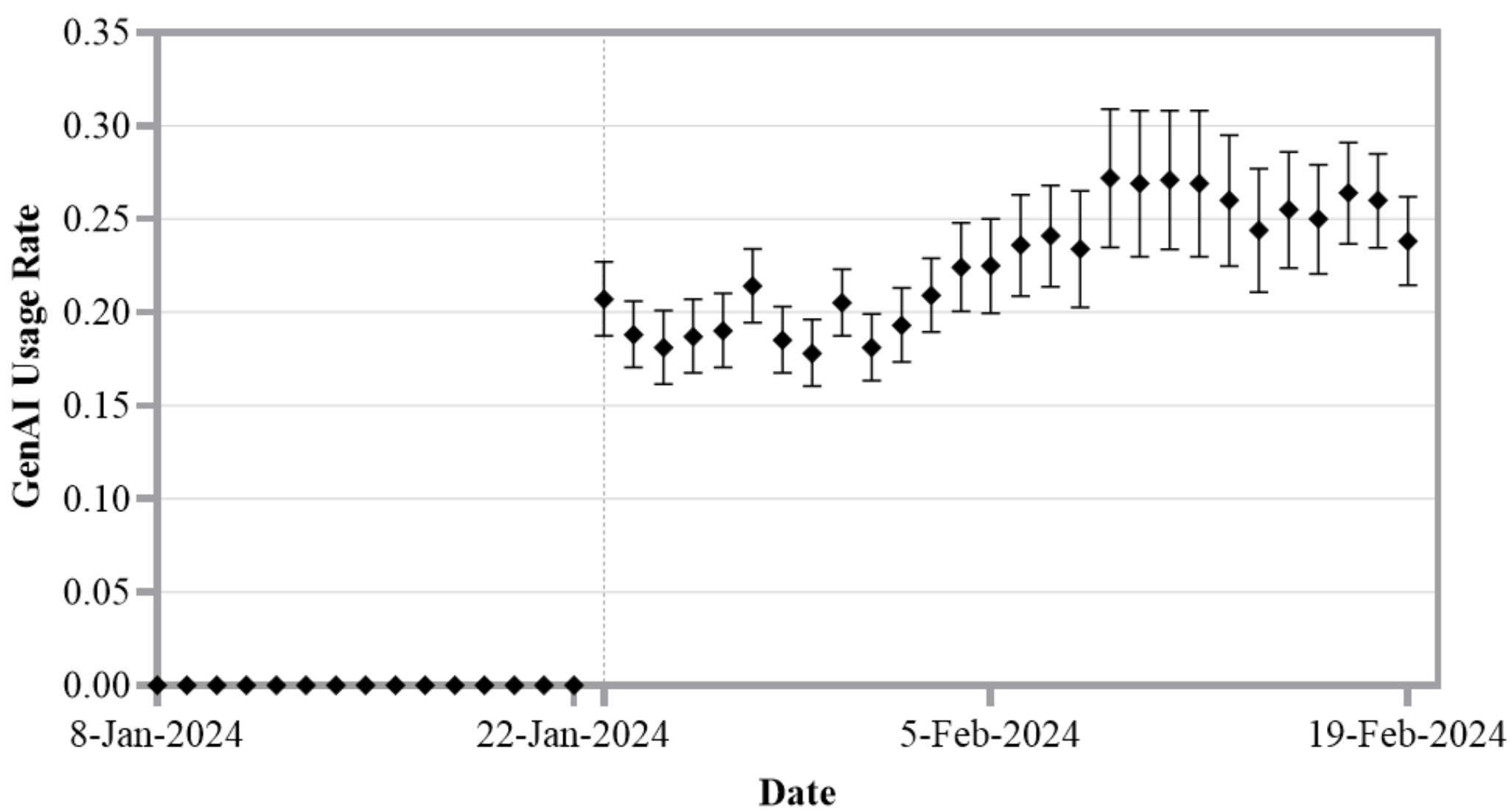

***Notes.*** Gen AI usage includes both complete and partial adherence. Average usage rates are calculated over all treated agents in a day. The error bars indicate 95% confidence intervals.

**Figure A8. Complete and Partial Adherence to Gen AI Over Time**

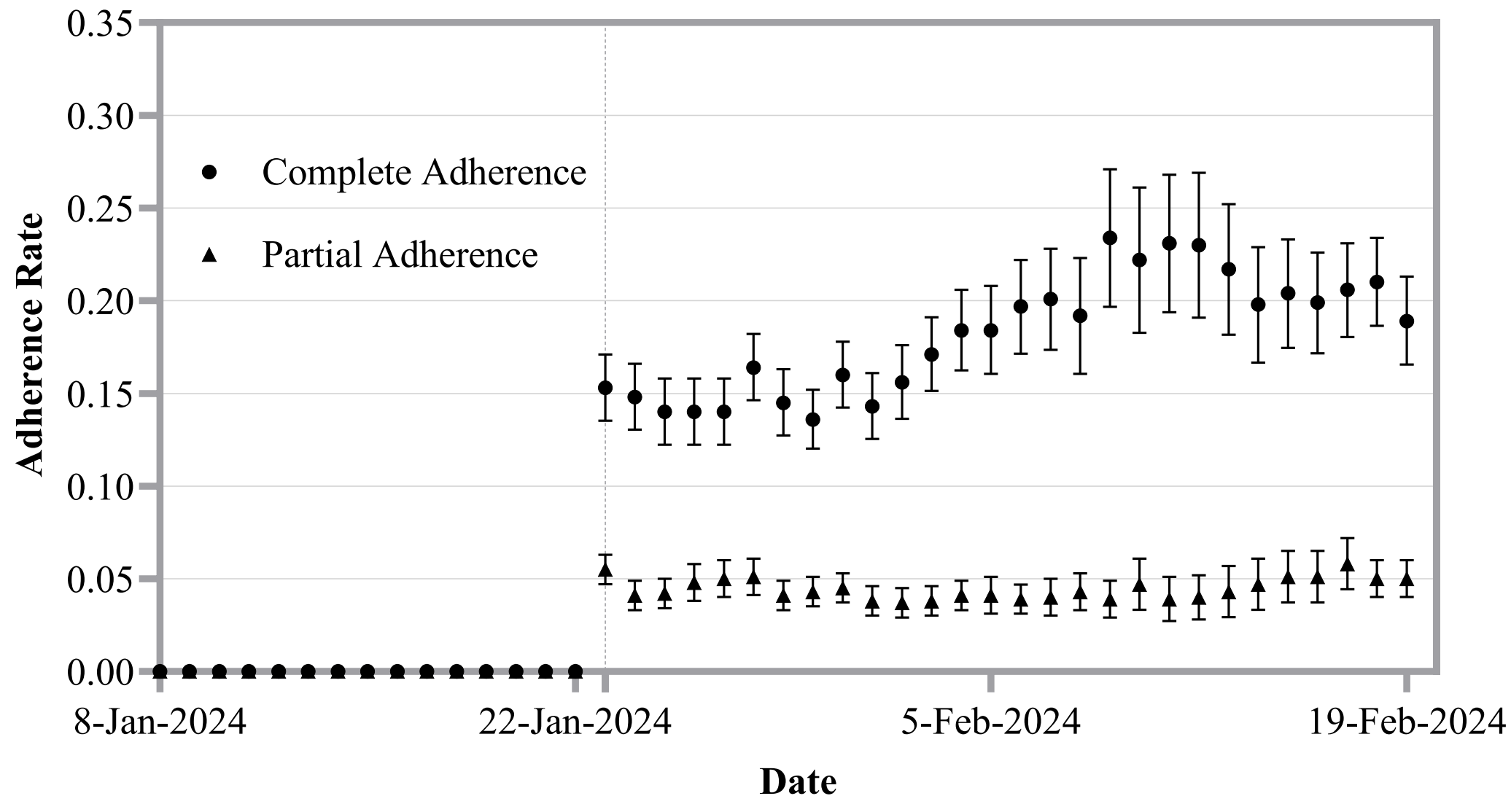

***Notes.*** Average adherence rates are calculated over all treated agents in a day. The error bars indicate 95% confidence intervals.

**Figure A9. Model-Free Evidence**

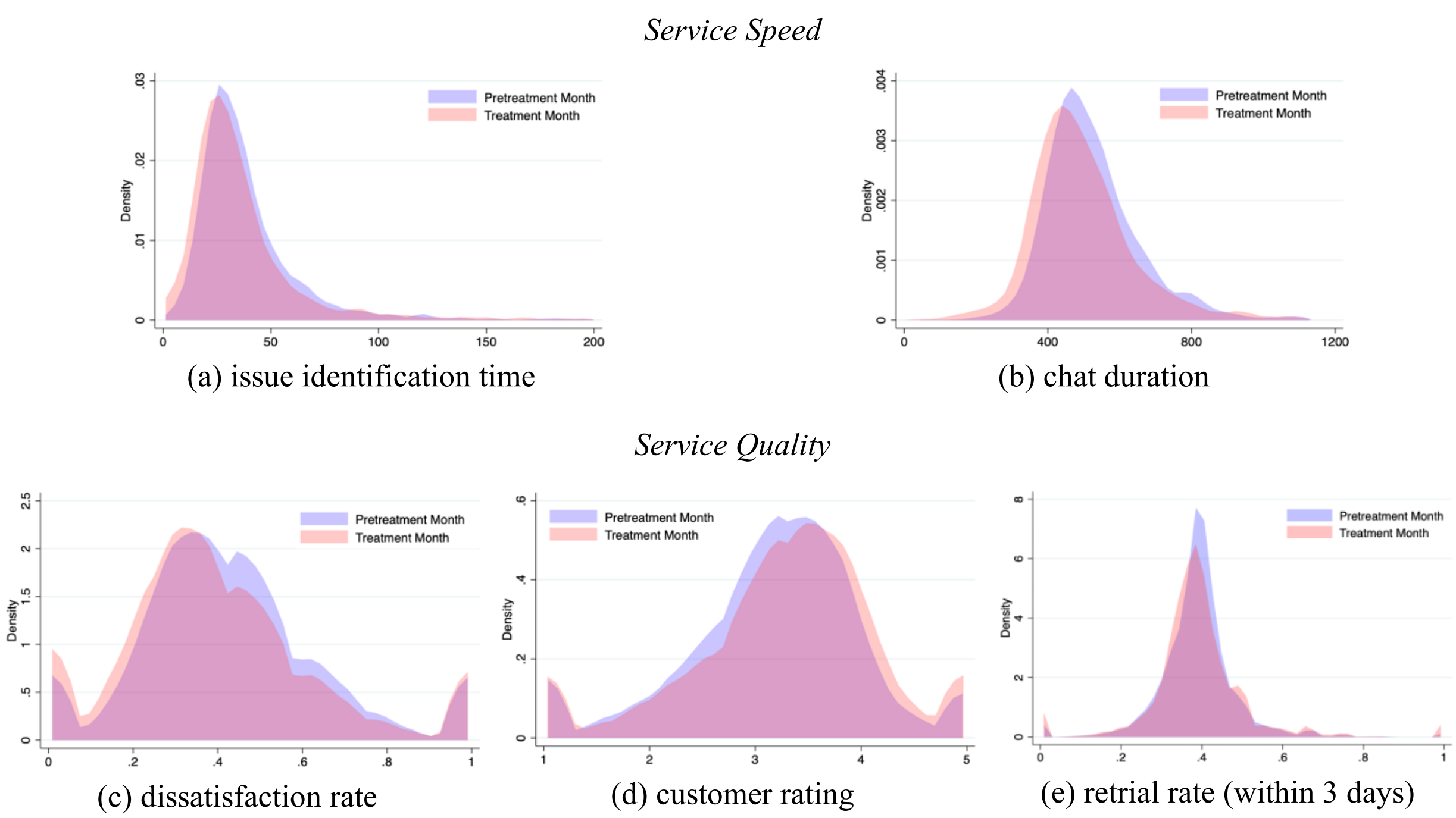

(a) issue identification time (b) chat duration

(c) dissatisfaction rate (d) customer rating (e) retrial rate (within 3 days)

***Notes.*** These figures illustrate the distributional changes in treated agents' performance between the pretreatment and treatment periods. The x-axis represents average performance at the agent-month level, while the y-axis indicates the corresponding density.

## Appendix B. Tables

## Table A1. Balance Check

| | treated<br>717,650 | control<br>717,270 | p-value |
|---|---|---|---|
| **Chat Information** | | | |
| issue identification time (seconds) | 34.88 | 36.40 | 0.120 |
| | (0.67) | (0.71) | |
| chat duration (seconds) | 474.93 | 475.69 | 0.815 |
| | (2.31) | (2.18) | |
| if rated | 0.15 | 0.15 | 0.526 |
| | (0.00) | (0.00) | |
| agent response time (seconds) | 174.91 | 175.15 | 0.888 |
| | (1.20) | (1.14) | |
| customer response time (seconds) | 269.59 | 269.21 | 0.818 |
| | (1.20) | (1.12) | |
| number of agent messages | 11.90 | 11.83 | 0.497 |
| | (0.07) | (0.07) | |
| number of customer messages | 11.49 | 11.46 | 0.641 |
| | (0.04) | (0.04) | |
| number of customer pictures | 0.93 | 0.93 | 0.520 |
| | (1.77) | (1.79) | |
| total agent typing time (seconds) | 109.45 | 110.46 | 0.471 |
| | (1.03) | (0.94) | |
| concurrency | 2.46 | 2.45 | 0.313 |
| | (0.01) | (0.01) | |
| agent message length | 15.78 | 15.72 | 0.649 |
| | (0.10) | (0.10) | |
| agent lexical density | 0.50 | 0.50 | 0.291 |
| | (0.00) | (0.00) | |
| agent lexical diversity | 4.43 | 4.41 | 0.274 |
| | (0.01) | (0.01) | |
| customer message length | 8.92 | 8.93 | 0.711 |
| | (0.01) | (0.01) | |
| customer lexical density | 0.55 | 0.55 | 0.477 |
| | (0.00) | (0.00) | |
| customer lexical diversity | 3.77 | 3.77 | 0.684 |
| | (0.00) | (0.00) | |
| agent active time | 290.91 | 293.51 | 0.280 |
| | (1.70) | (1.71) | |
| agent shift-away time | 184.34 | 182.50 | 0.325 |
| | (1.45) | (1.18) | |
| customer retrial in 3 days | 0.39 | 0.39 | 0.988 |
| | (0.00) | (0.00) | |

## Table A1. Balance Check (continued)

| | treated<br>2,895 | control<br>3,045 | p-value |
|---|---|---|---|
| **Agent Information** | | | |
| tenure (days) | 124.96 | 122.77 | 0.118 |
| | (54.50) | (53.38) | |
| age | 31.76 | 31.74 | 0.930 |
| | (6.10) | (6.11) | |
| gender (M=1) | 0.18 | 0.17 | 0.281 |
| | (0.39) | (0.38) | |

| | treat<br>108,948 | control<br>108,306 | p-value |
|---|---|---|---|
| **Customer Rating Information** | | | |
| rating | 3.46 | 3.46 | 0.958 |
| | (0.01) | (0.01) | |
| if dissatisfied | 0.35 | 0.35 | 0.970 |
| | (0.00) | (0.00) | |

## Table A2. Antecedents of Gen AI Usage

| | GenAI_Usage | | | | |
|---|---|---|---|---|---|
| | (1) | (2) | (3) | (4) | (5) |
| Male | -0.007 | -0.009 | -0.019 | -0.023 | -0.019 |
| | (0.017) | (0.017) | (0.017) | (0.017) | (0.017) |
| Age | -0.002** | -0.002** | -0.001 | -0.002 | -0.002* |
| | (0.001) | (0.001) | (0.001) | (0.001) | (0.001) |
| ln(Tenure) | | -0.038*** | -0.051*** | -0.046** | -0.042** |
| | | (0.013) | (0.018) | (0.018) | (0.018) |
| Pretreatment Customer Rating | | | -0.053*** | -0.038*** | -0.035*** |
| | | | (0.008) | (0.008) | (0.008) |
| Pretreatment Concurrency | | | | -0.114*** | -0.107*** |
| | | | | (0.013) | (0.013) |
| Pretreatment Chat Duration | | | | | 0.000** |
| | | | | | (0.000) |
| Constant | 0.412*** | 0.588*** | 0.773*** | 0.950*** | 0.857*** |
| | (0.035) | (0.069) | (0.095) | (0.096) | (0.103) |
| Observations | 2,856 | 2,856 | 2,482 | 2,482 | 2,482 |

***Notes.*** *p<0.1; **p<0.05; ***p<0.01. Agent tenure (days) is log-transformed. Standard errors are in parentheses.

## Table A3. Antecedents of Gen AI Usage Across Chat Characteristics

| | if adhere to gen AI suggestion | | | |
|---|---|---|---|---|
| | (1) | (2) | (3) | (4) |
| If refund issue | 0.015*** | 0.016*** | 0.016*** | 0.017*** |
| | (0.003) | (0.003) | (0.003) | (0.003) |
| If shipping fee issue | 0.019*** | 0.019*** | 0.019*** | 0.019*** |
| | (0.003) | (0.003) | (0.003) | (0.003) |
| If exchange issue | 0.013*** | 0.014*** | 0.014*** | 0.014*** |
| | (0.004) | (0.004) | (0.004) | (0.004) |
| If reshipment issue | 0.020*** | 0.021*** | 0.021*** | 0.021*** |
| | (0.004) | (0.004) | (0.004) | (0.004) |
| If repair issue | 0.022*** | 0.024*** | 0.024*** | 0.023*** |
| | (0.005) | (0.005) | (0.005) | (0.005) |
| If first contact | | 0.008*** | 0.008*** | 0.008*** |
| | | (0.001) | (0.001) | (0.001) |
| Concurrency | | | -0.006*** | -0.006*** |
| | | | (0.001) | (0.001) |
| Customer tenure | | | | 0.001*** |
| | | | | (0.000) |
| Agent Fixed Effects | Y | Y | Y | Y |
| Day Fixed Effects | Y | Y | Y | Y |
| Observations | 543,591 | 543,591 | 543,531 | 542,544 |

***Notes.*** *p<0.1; **p<0.05; ***p<0.01. Robust standard errors are clustered by agent.

## Table A4. Placebo Test: Gen AI's ITT Impact on "Never-Takers" in the Treatment Group

| | **Service Speed** | | **Service Quality** | | |
|---|---|---|---|---|---|
| | ln(identification time) | ln(chat duration) | dissatisfaction rate | customer rating | retrial rate |
| | (1) | (2) | (3) | (4) | (5) |
| Treat | 0.010 | -0.008 | -0.011 | 0.039 | -0.000 |
| | (0.019) | (0.008) | (0.008) | (0.030) | (0.003) |
| Agent Fixed Effects | Y | Y | Y | Y | Y |
| Day Fixed Effects | Y | Y | Y | Y | Y |
| Time-Variant Controls | Y | Y | Y | Y | Y |
| Observations | 1,415,461 | 1,480,420 | 223,601 | 223,601 | 1,480,420 |

***Notes.*** This table shows the estimated treatment effects on "never-takers" in the treatment group. *p<0.1; **p<0.05; ***p<0.01. Robust standard errors are in parentheses.

## Table A5. Gen AI's Effects on the Customer's Likelihood to Rate a Chat

| | if rated |
|---|---|
| | (1) |
| Treat | 0.0001 |
| | (0.001) |
| Agent Fixed Effects | Y |
| Day Fixed Effects | Y |
| Time-Variant Controls | Y |
| Observations | 2,564,221 |

***Notes.*** *p<0.1; **p<0.05; ***p<0.01. Robust standard errors are in parentheses.

## Table A6. Gen AI's LATE Effects on Agent-Customer Chat Textual Characteristics

**Panel A:** ***Issue Identification and Solution Proposal Stage***

| | Agent | | | Customer | | |
|---|---|---|---|---|---|---|
| | message length | lexical density | lexical diversity | message length | lexical density | lexical diversity |
| | (1) | (2) | (3) | (4) | (5) | (6) |
| GenAI_Usage | 7.674*** | 0.052*** | 1.713*** | -1.339*** | -0.059*** | -0.254*** |
| | (0.237) | (0.005) | (0.054) | (0.229) | (0.004) | (0.031) |
| Agent Fixed Effects | Y | Y | Y | Y | Y | Y |
| Day Fixed Effects | Y | Y | Y | Y | Y | Y |
| Time-Variant Controls | Y | Y | Y | Y | Y | Y |
| Observations | 126,255 | 126,255 | 126,255 | 126,271 | 126,264 | 126,271 |

**Panel B:** ***Resolution Stage***

| | Agent | | | Customer | | |
|---|---|---|---|---|---|---|
| | message length | lexical density | lexical diversity | message length | lexical density | lexical diversity |
| | (1) | (2) | (3) | (4) | (5) | (6) |
| GenAI_Usage | -0.428 | -0.001 | -0.045 | -0.361* | 0.010 | -0.087** |
| | (0.329) | (0.004) | (0.043) | (0.193) | (0.006) | (0.041) |
| Agent Fixed Effects | Y | Y | Y | Y | Y | Y |
| Day Fixed Effects | Y | Y | Y | Y | Y | Y |
| Time-Variant Controls | Y | Y | Y | Y | Y | Y |
| Observations | 118,458 | 118,458 | 118,458 | 122,196 | 122,179 | 122,196 |

**Panel C:** ***Confirmation Stage***

| | Agent | | | Customer | | |
|---|---|---|---|---|---|---|
| | message length | lexical density | lexical diversity | message length | lexical density | lexical diversity |
| | (1) | (2) | (3) | (4) | (5) | (6) |
| GenAI_Usage | 0.139 | 0.001 | -0.029 | -0.339* | -0.004 | -0.093** |
| | (0.320) | (0.003) | (0.033) | (0.185) | (0.005) | (0.040) |
| Agent Fixed Effects | Y | Y | Y | Y | Y | Y |
| Day Fixed Effects | Y | Y | Y | Y | Y | Y |
| Time-Variant Controls | Y | Y | Y | Y | Y | Y |
| Observations | 124,503 | 124,503 | 124,503 | 122,744 | 122,692 | 122,744 |

***Notes.*** *p<0.1; **p<0.05; ***p<0.01. Robust standard errors are clustered at the agent level.

## Table A7. Gen AI Adherence Across Worker Performance Quintiles: Descriptive Statistics

| | if adhere to gen AI suggestion |
|---|---|
| Q1 Agents | 0.393 |
| Q2 Agents | 0.253 |
| Q3 Agents | 0.219 |
| Q4 Agents | 0.180 |
| Q5 Agents | 0.183 |

## Table A8. Average Customer Ratings of Treated Agents Across Performance Quintiles

| | average customer rating | |
|---|---|---|
| | pretreatment | treatment |
| Q1 Agents | 2.184 | 2.981 |
| Q2 Agents | 2.881 | 3.145 |
| Q3 Agents | 3.271 | 3.406 |
| Q4 Agents | 3.637 | 3.666 |
| Q5 Agents | 4.008 | 3.846 |

## Table A9. Distribution of Customer Issue Categories for Treated Agents Across Performance Quintiles

| | pretreatment | | | | | posttreatment | | | | |
|---|---|---|---|---|---|---|---|---|---|---|
| category | Q1 | Q2 | Q3 | Q4 | Q5 | Q1 | Q2 | Q3 | Q4 | Q5 |
| refund | 77.82% | 77.76% | 77.52% | 77.30% | 77.00% | 77.80% | 77.04% | 76.93% | 76.89% | 77.09% |
| shipping fee | 14.57% | 14.80% | 14.85% | 14.96% | 14.78% | 12.91% | 13.91% | 13.87% | 13.50% | 13.11% |
| exchange | 2.73% | 2.81% | 2.83% | 2.82% | 3.13% | 3.11% | 3.02% | 2.93% | 3.10% | 3.18% |
| reshipment | 1.68% | 1.75% | 1.86% | 1.97% | 1.98% | 2.05% | 2.35% | 2.45% | 2.55% | 2.61% |
| repair | 1.08% | 1.07% | 1.16% | 1.13% | 1.13% | 1.18% | 1.17% | 1.20% | 1.17% | 1.12% |
| others | 2.12% | 1.81% | 1.78% | 1.82% | 1.98% | 2.95% | 2.51% | 2.62% | 2.79% | 2.89% |

## Table A10. Issue Identification Match Rate Across Performance Quintiles for Chats Handled by Treated Agents: Regression Results

| | issue identification match rate | | | |
|---|---|---|---|---|
| | full sample | first contact | shipping fee | reshipment |
| | (1) | (2) | (3) | (4) |
| Q2 Agents | 0.001 | 0.003 | -0.012 | 0.003 |
| | (0.004) | (0.004) | (0.010) | (0.025) |
| Q3 Agents | 0.000 | 0.000 | -0.010 | -0.011 |
| | (0.004) | (0.004) | (0.010) | (0.023) |
| Q4 Agents | -0.003 | -0.003 | -0.009 | -0.018 |
| | (0.004) | (0.004) | (0.010) | (0.023) |
| Q5 Agents | -0.006 | -0.006 | -0.016 | -0.019 |
| | (0.004) | (0.005) | (0.010) | (0.024) |
| Day Fixed Effects | Y | Y | Y | Y |
| Time-Variant Controls | Y | Y | Y | Y |
| Observations | 529,546 | 394,991 | 72,096 | 13,246 |
| R-squared | 0.003 | 0.002 | 0.031 | 0.009 |

***Notes.*** The analysis is conducted at the chat level. The sample consists of chats handled by treated agents during the treatment period. Q1 agents constitute the omitted reference group. *p<0.1; **p<0.05; ***p<0.01. Robust standard errors are in parentheses.

## Table A11. Gen AI Adherence Across Worker Performance Quintiles: Regression Results

| | if adhere to gen AI suggestion | | | |
|---|---|---|---|---|
| | full sample | first contact | shipping fee | reshipment |
| | (1) | (2) | (3) | (4) |
| Q2 Agents | -0.122*** | -0.115*** | -0.111*** | -0.106*** |
| | (0.030) | (0.030) | (0.031) | (0.039) |
| Q3 Agents | -0.143*** | -0.138*** | -0.132*** | -0.133*** |
| | (0.029) | (0.029) | (0.030) | (0.036) |
| Q4 Agents | -0.172*** | -0.167*** | -0.169*** | -0.156*** |
| | (0.029) | (0.029) | (0.029) | (0.036) |
| Q5 Agents | -0.163*** | -0.159*** | -0.163*** | -0.167*** |
| | (0.030) | (0.030) | (0.030) | (0.036) |
| Day Fixed Effects | Y | Y | Y | Y |
| Time-Variant Controls | Y | Y | Y | Y |
| Observations | 529,546 | 394,991 | 72,096 | 13,264 |
| R-squared | 0.026 | 0.023 | 0.027 | 0.023 |

***Notes.*** The analysis is conducted at the chat level. The sample consists of chats handled by treated agents during the treatment period. Q1 agents constitute the omitted reference group. *p<0.1; **p<0.05; ***p<0.01. Robust standard errors are in parentheses.

## Table A12. Destination of Shift-Away from the Focal Chat Across Non-Focal Chat Stages

| | ***Shift-Away Destination Stage*** | | |
|---|---|---|---|
| | issue identification and solution proposal | resolution | confirmation |
| Q1 agents | 60.1% | 21.6% | 18.3% |
| Q2 agents | 52.2% | 25.6% | 22.2% |
| Q3 agents | 49.1% | 26.5% | 24.5% |
| Q4 agents | 49.1% | 27.8% | 23.1% |
| Q5 agents | 50.1% | 27.1% | 22.8% |

***Notes.*** This table reports the distribution of non-focal chat stages at the moment an agent shifts away from a focal chat to attend to another ongoing chat.

## Table A13. Gen AI's LATE Effects (with Agent-Hour Level Data)

| | **Service Speed** | | **Service Quality** | | |
|---|---|---|---|---|---|
| | ln(identification time) | ln(chat duration) | dissatisfaction rate | customer rating | retrial rate |
| | (1) | (2) | (3) | (4) | (5) |
| GenAI_Usage | -0.374*** | -0.075*** | -0.049** | 0.179** | -0.004 |
| | (0.053) | (0.021) | (0.019) | (0.073) | (0.008) |
| Agent Fixed Effects | Y | Y | Y | Y | Y |
| Day Fixed Effects | Y | Y | Y | Y | Y |
| Hour Fixed Effects | Y | Y | Y | Y | Y |
| Time-Variant Controls | Y | Y | Y | Y | Y |
| Observations | 579,117 | 588,066 | 265,208 | 265,208 | 588,066 |

***Notes.*** *p<0.1; **p<0.05; ***p<0.01. Robust standard errors are clustered by agent.

## Table A14. Gen AI's LATE Heterogeneous Effects (with Agent-Hour Level Data)

| | Service Speed | | Service Quality | | |
|---|---|---|---|---|---|
| | ln(identification time) | ln(chat duration) | dissatisfaction rate | customer rating | retrial rate |
| | (1) | (2) | (3) | (4) | (5) |
| GenAI_Usage | -0.399*** | -0.213*** | -0.553*** | 2.088*** | -0.004 |
| | (0.081) | (0.034) | (0.049) | (0.190) | (0.016) |
| GenAI_Usage × Q2 Agents | -0.115 | 0.099** | 0.377*** | -1.424*** | 0.004 |
| | (0.110) | (0.046) | (0.058) | (0.224) | (0.021) |
| GenAI_Usage × Q3 Agents | -0.110 | 0.065 | 0.458*** | -1.734*** | -0.011 |
| | (0.110) | (0.044) | (0.055) | (0.213) | (0.019) |
| GenAI_Usage × Q4 Agents | 0.134 | 0.220*** | 0.557*** | -2.078*** | -0.015 |
| | (0.113) | (0.047) | (0.055) | (0.214) | (0.020) |
| GenAI_Usage × Q5 Agents | 0.238** | 0.242*** | 0.815*** | -3.163*** | 0.042* |
| | (0.118) | (0.055) | (0.070) | (0.270) | (0.023) |
| Impact on Q1 Agents | -0.399*** | -0.213*** | -0.553*** | 2.088*** | -0.004 |
| Impact on Q2 Agents | -0.514*** | -0.113*** | -0.176*** | 0.664*** | 0.000 |
| Impact on Q3 Agents | -0.509*** | -0.148*** | -0.095*** | 0.354*** | -0.015 |
| Impact on Q4 Agents | -0.265*** | 0.007 | 0.005 | 0.011 | -0.018 |
| Impact on Q5 Agents | -0.161* | 0.029 | 0.263*** | -1.075*** | 0.038** |
| Agent Fixed Effects | Y | Y | Y | Y | Y |
| Day Fixed Effects | Y | Y | Y | Y | Y |
| Hour Fixed Effects | Y | Y | Y | Y | Y |
| Time-Variant Controls | Y | Y | Y | Y | Y |
| Observations | 569,173 | 577,818 | 262,345 | 262,345 | 577,818 |

***Notes.*** *The data are at the agent-hour level.* *p<0.1; **p<0.05; ***p<0.01. Robust standard errors are clustered by agent.

## Appendix C. Definitions of Agent-Customer Interaction Measures

This appendix defines the process-level variables used to study how gen AI changes agent-customer interactions. The measures are grouped into four categories: (1) agent-customer chat interaction measures; (2) textual characteristics of agent and customer messages; (3) agent-customer message ratios across chat stages; and (4) agent shift-away measures.

### C1. Agent-Customer Chat Interaction Measures

We construct interaction measures from session logs, message timestamps, message content, and message type labels. These measures capture how quickly agents and customers respond, how much each party communicates, and how much effort customers provide during the interaction. All time-related measures are recorded in seconds.

**Figure A10. Agent-Customer Chat Process Measures for a Hypothetical Chat Session**

**Agent-Customer Chat Process Measures**

| Agent | | | Customer | | | |
|---|---|---|---|---|---|---|
| Response Time | # of Messages | Typing Time | Response Time | # of Messages | # Short Affirmative Replies | # of Pictures |
| 6.5 minutes | 5 | 0.5 minutes | 2.5 minutes | 2 | 2 | 0 |

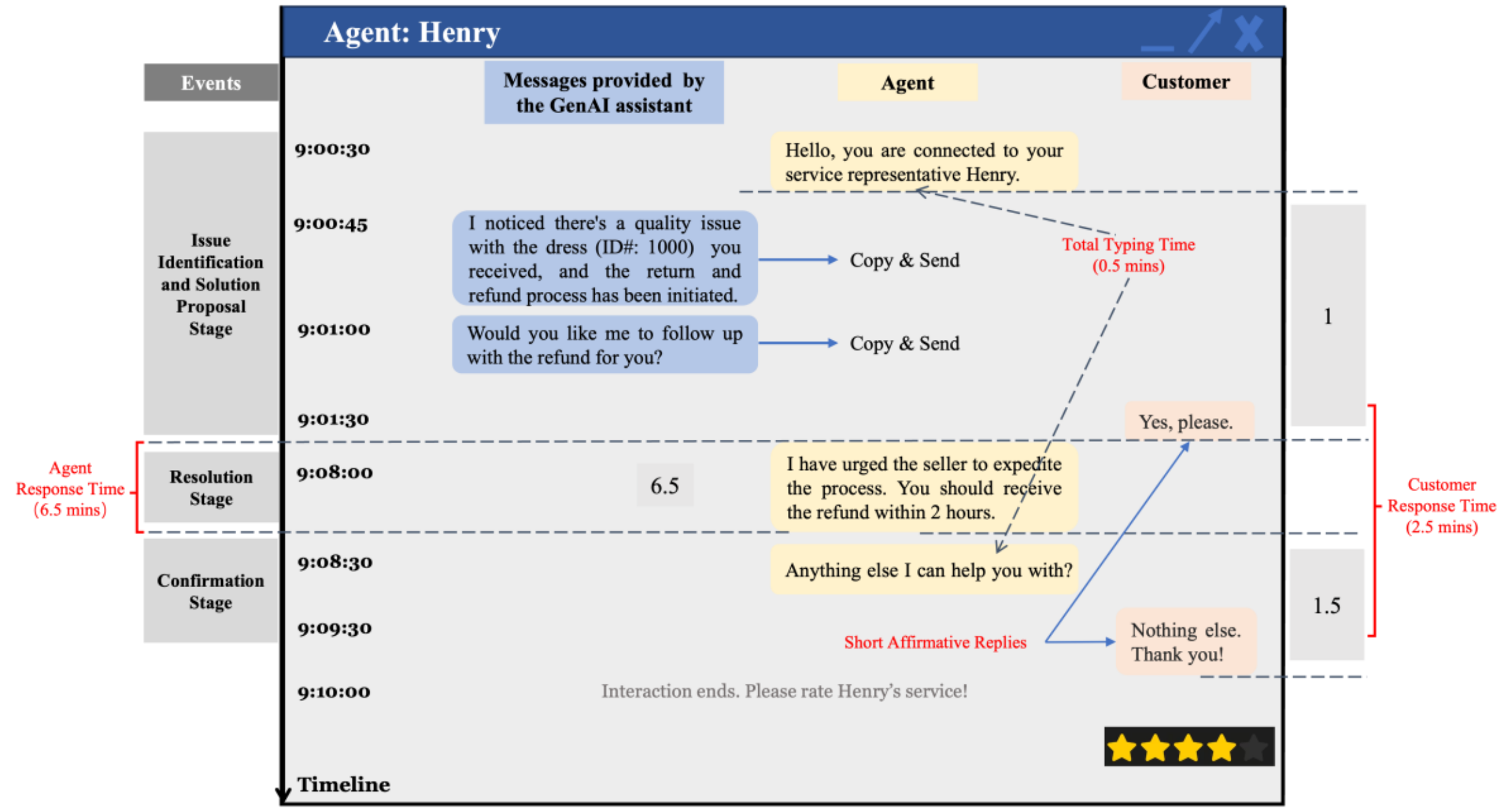

*Agent response time* and *customer response time* capture how quickly each party replies after the other party initiates a response round. A response round is a sequence in which one party sends one or more consecutive messages and the other party subsequently replies. Within each round, response time is measured as the elapsed time between the first message from the initiating party and the first subsequent reply from the responding party. For a given chat or chat stage, we sum response time across all applicable response rounds. If a party does not reply within a stage, the corresponding stage-specific response time is coded as missing for that party-stage observation. In the stage-specific analyses, response-time measures are computed within the chat-stage boundaries defined in Appendix C3. We reconstruct response time measures from message timestamps for each online service chat.

The *number of agent messages* and the *number of customer messages* refer to the total number of messages sent by each party during a chat or chat stage. These measures capture the intensity of communication. *Agent typing time* is obtained from session logs, which record the amount of time the agent's cursor remains active in the message input box. We use this measure as a proxy for agent effort in composing responses.

The *number of customer pictures* is defined as the count of image-type messages sent by the customer, as identified by message type labels. Customer picture uploads are interpreted as higher-effort diagnostic inputs because they typically require customers to provide visual evidence or documentation.

The *number of customer short affirmative replies* captures brief customer messages that signal confirmation, acknowledgment, gratitude, or agreement without providing substantive new diagnostic information. To construct this measure, we focus on customer messages with no more than six Chinese characters, review the most frequent short messages in the data, and include common low-effort replies. Examples include confirmations (e.g., "yes," "okay," "sure"), acknowledgments (e.g., "received," "got it"), gratitude expressions (e.g., "thank you," "thanks"), and affirmative responses (e.g., "done," "agreed").

Figure A10 illustrates these agent-customer interaction measures using a hypothetical chat session. For the agent, there is one response round in which the agent message "I have urged ... within 2 hours" at 9:08:00 is preceded by the customer's message "Yes, please." at 9:01:30. This round yields an agent response time of 6.5 minutes. For the customer, there are two agent-initiated response rounds: the agent message at 9:00:30 is followed by the customer's first subsequent reply at 9:01:30, and the agent message at 9:08:00 is followed by the customer's first subsequent reply at 9:09:30. These two rounds yield a total customer response time of 2.5 minutes. The customer sends two messages, both of which are short affirmative replies. The agent sends five messages and spends a total of 0.5 minutes typing them.

### C2. Agent-Customer Chat Textual Characteristics Measures

To quantify the textual characteristics of messages exchanged during a chat, we construct three textual measures adapted from Read (2000): message length, lexical density, and lexical diversity. Each measure is computed separately for agents and customers. System-generated messages and non-text messages are excluded from the textual analyses.

*Message length* is measured by the average number of Chinese characters per message, reflecting the degree to which a message is elaborate. *Lexical density* is defined as the ratio of content words (such as nouns, verbs, adjectives) to the total number of words in a message, capturing the informational richness of the communication. *Lexical diversity* refers to entropy, calculated as $-\sum p \log_2 p$, where $p$ represents the frequency of each content word. Higher entropy values indicate greater variation in word usage, and thus more diverse language production.

### C3. Agent-Customer Message Ratio Across Chat Stages

To measure the relative intensity of agent-customer communication, we construct the *agent-customer message ratio*, defined as the number of agent messages divided by the number of customer messages. We compute this ratio for the full chat and separately for each chat stage.

To define these stages, we use Alibaba's standardized operating procedures (SOPs) to partition each chat into three sequential stages. SOPs are structured decision paths that guide agents from issue diagnosis to predefined responses and solutions, and the system records the timing of SOP-based messages. First, the *issue identification and solution proposal stage* is defined as the period from chat start to the first SOP-based message, when the agent diagnoses the issue and proposes a solution. Second, the *resolution stage* covers the SOP-based implementation of the solution. Third, the *confirmation stage* includes the remaining exchange after resolution, when the agent confirms whether the customer's issue has been addressed. These stage definitions allow us to measure process-level outcomes for the full chat and for specific stages of the chat. Figure A10 illustrates the stage boundaries using a hypothetical chat.

### C4. Agent Shift-Away Measures

We use the timestamps of agents' messages across concurrent chats to infer when an agent shifts away from the focal chat. After an agent sends a message in the focal chat, we identify a shift-away event if the agent's next observed message is sent in another concurrent chat. Activities such as checking customer history or order details are not observed as chat messages and are therefore not classified as shift-away events. The shift-away interval begins after the focal-chat message and ends when the agent returns to the focal chat. We interpret this interval as time during which the agent is attending to other concurrent chats rather than actively messaging in the focal chat.

Based on this approach, we construct three measures: *active time*, *shift-away time*, and *share of shift-away time. Active time* captures the cumulative duration during which an agent actively engages with the focal chat (i.e., not attending to other concurrent chats). *Shift-away time* captures the cumulative duration during which the agent is observed in another concurrent chat and is calculated as the difference between total chat duration and active time. *Share of shift-away time* represents the proportion of time the agent spends away from the focal chat, calculated as shift-away time divided by chat duration.

**Figure A11. Illustration of Agent Task-Switching for a Hypothetical Chat Session**

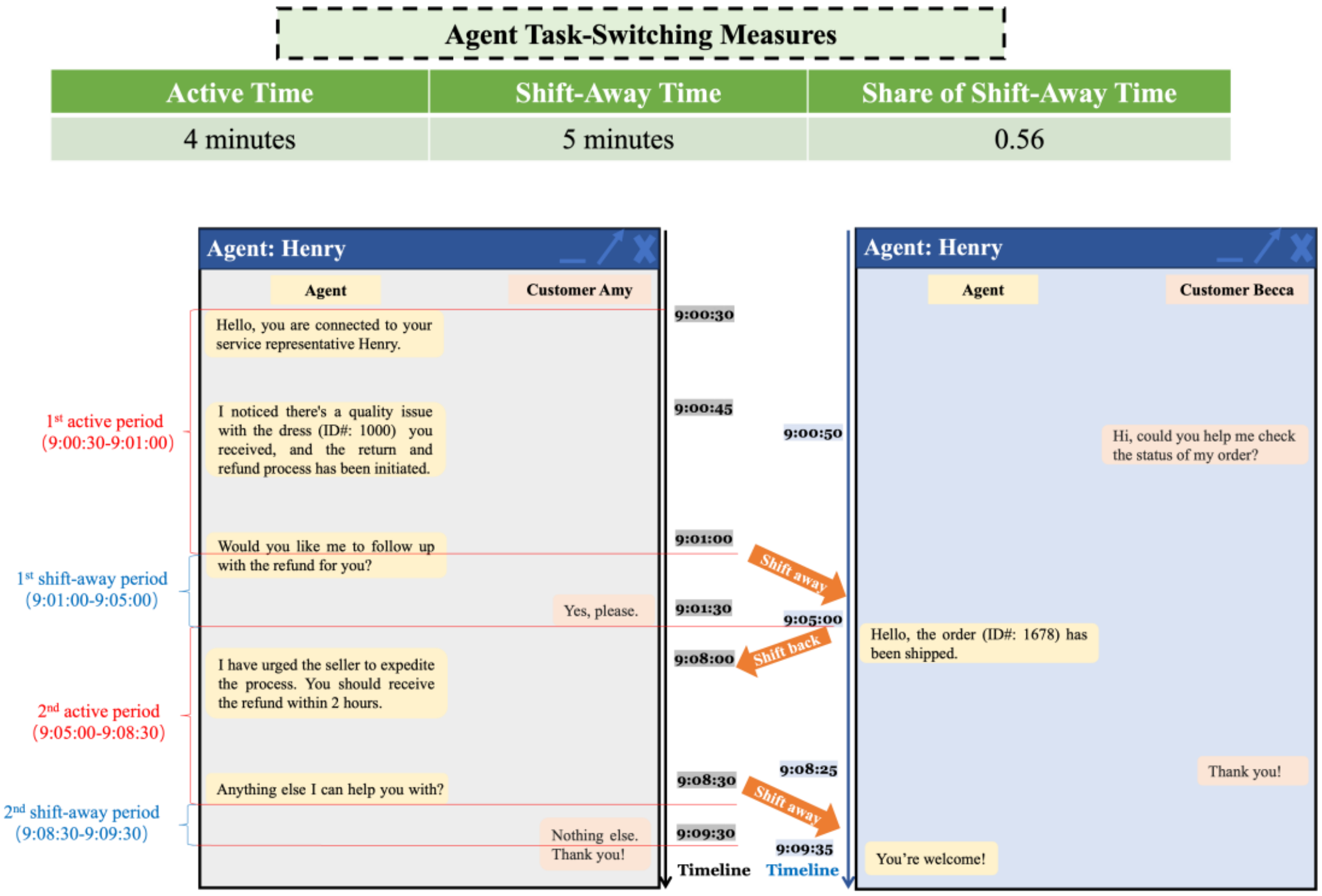

Figure A11 illustrates the calculation of these measures using a hypothetical chat session in which the agent manages two chats simultaneously. Based on overlapping message timelines, the first shift-away occurs after the agent sends the message "Would you…for you?" (9:01:00), followed by a message in another chat. The second shift-away occurs after the message "Anything else I can help you with?" (9:08:30), again followed by activity in another session. The two active periods span from 9:00:30 to 9:01:00 and from 9:05:00 to 9:08:30, yielding a total active time of four minutes. Correspondingly, the two shift-away periods span from 9:01:00 to 9:05:00 and from 9:08:30 to 9:09:30, totaling five minutes. Given a total chat duration of nine minutes, the agent spends five minutes away from the focal chat. The resulting share of shift-away time is 5/9, or 0.56.

## Appendix D. Construction of the Issue Identification Matching

We construct a chat-level issue identification match indicator to assess whether the issue category inferred from the gen AI suggestion aligns with the issue category recorded by the human agent at the end of the chat. The indicator equals one when the inferred gen AI issue category matches the human-recorded issue category and zero otherwise. According to Alibaba's SOPs, agents are required to select an issue category upon chat completion. As shown in Table A9, after-sales inquiries are concentrated in five issue categories: refund, shipping fee, exchange, reshipment, and repair, which together account for 97.27% of all customer issues.

We infer the issue category identified by gen AI using a keyword-based mapping procedure. The mapping is applied first to the proposed solution text. If no issue category is identified from the proposed solution text, we then apply the mapping to the diagnosis text. Table A15 reports the keyword dictionary used to map AI suggestions to the five issue categories. The dictionary was constructed based on Alibaba's issue category definitions and frequently observed expressions in the gen AI suggestion text.

Some AI suggestions are generic or non-diagnostic, such as “How may I help you?” or “after-sales support.” These suggestions cannot be mapped to a specific issue category and are listed in Table A16. Generic suggestions are coded as mismatches except when they appear during the customer’s first contact. Overall, 92.8% of AI suggestions can be assigned either to one of the five issue categories or to a generic suggestion type. If no issue category can be identified, the suggestion is coded as a mismatch, making the measure conservative.

**Table A15. Keyword Dictionary for Mapping AI Suggestions to Issue Categories**

| Issue category | Keywords | Share of AI suggestions |
|---|---|---|
| Refund | return; refund, refund destination / status | 54.18% |
| Shipping fee | shipping fee | 10.42% |
| Exchange | exchange | 2.02% |
| Reshipment | re-ship; re-send | 1.82% |
| Repair | repair; warranty | 0.43% |

**Table A16. Generic Suggestion Types (Excluded from Category Mapping)**

| Suggestion type | Keywords | Share of AI suggestions |
|---|---|---|
| Generic help | How may I help you? | 5.19% |
| Generic support | after-sales support | 6.36% |
| Logistics | address; pickup; courier | 3.27% |
| Shipping progress | shipped; sent out | 4.87% |
| Platform escalation/status | current situation; platform is handling | 0.77% |
| Expedite | expedite / urge | 3.47% |
| Unmatched | No matched tokens | 7.20% |

## Appendix E. Online Survey Results

Alibaba internally conducted an online survey five months after the experiment to assess treated agents’ perceptions of the gen AI assistant. The survey was distributed via the enterprise collaboration platform *Ali Ding* and was fully anonymous to encourage candid responses. A total of 834 valid responses were collected. The survey consisted of two parts: Part A focused on agents’ experiences with the gen AI assistant, and Part B collected agents’ self-assessments of their performance.

Based on Part B, 63.4% of respondents worked for less than one year, 17.5% for one to two years, and 19.1% for more than two years. Agents’ skill level was proxied by their performance rank, ranging from one (lowest) to five (highest), which positively correlates with pretreatment service performance. As shown in Table A19, self-assessed performance generally increases with rank, with slight deviation at level one due to the small sample size. Among all respondents, 20 agents were ranked level one, 171 level two, 314 level three, 236 level four, and 93 level five. In the following analysis, we focus on high performers (level 5 agents) and low performers (levels 1 and 2)[12].

In Part A, agents were first asked, "How frequently do you follow the gen AI assistant's suggestions?" Among respondents, 47.1% selected Sometimes follow, 41.6% Often follow, 6.7% Never follow, and 4.6% Always follow. These responses broadly align with the average gen AI usage rate of 28.7% observed during the experiment and the presence of both "always-followers" and "never-takers" throughout the experimental period.

Next, agents were asked, "Under what conditions do you tend to follow the gen AI assistant?" with four possible options: (1) during peak hours or periods of high workload, (2) when workload is relatively light, (3) when handling complex cases requiring customer history review, and (4) when dealing with simple cases. The most frequently selected condition was "during peak hours or periods of high workload" (66.9%), followed by "when handling complex cases" (52.4%). A smaller share of respondents selected "when workload is relatively light" (8.8%) or "when dealing with simple cases" (28.2%).

Agents were also asked to rate several statements about their experience with the gen AI assistant on a five-point Likert scale (1 = *Strongly Disagree*, 5 = *Strongly Agree*). The detailed distributions by performance level are presented in Tables A17 and A18. For perceived suggestion quality (A1), high performers reported a lower mean score (3.72) than low performers (3.83). A similar pattern emerged for perceived helpfulness (A2): 71.2% of low performers agreed that the gen AI assistant was helpful, compared with 62.4% of high performers. Regarding specific performance measures, low performers reported greater perceived improvements than high performers in both service speed (77.5% vs. 67.7%) and service quality (66.5% vs. 61.3%). Across all respondents, the assistant was viewed as more effective in improving service speed than quality.

Among the detailed benefits (A5–A11), *responsiveness* (A8) received the highest average rating, suggesting that agents found the gen AI assistant most useful for accelerating replies. *Case summarization* (A6) ranked second. For all gen AI's benefits, high performers consistently provided lower ratings than low performers.

The final question, "Do you have any suggestions or comments on the gen AI assistant?", was open-ended. Among the 35 high performers who provided non-empty responses, 51% expressed concerns regarding accuracy, while 26% offered suggestions for improvement. These qualitative comments are consistent with the quantitative results: high performers generally used the gen AI assistant with greater caution, frequently verifying its outputs before use, which likely contributed to their lower adoption rate of the tool.

[12] All observations continue to hold if we include both level 4 and 5 agents in the high-performer group.

## Table A17. Survey Responses of High-Performing Agents on Their Experience with the Gen AI Assistant

| Do you agree with the following statements? Please indicate how much you agree or disagree with the following statements.<br>No. | Statement | Strongly Disagree | Disagree | Neutral | Agree | Strongly Agree |
|---|---|---|---|---|---|---|
| A1 | The quality of suggestions generated by the gen AI assistant is good. | 2.2% | 7.5% | 30.1% | 36.6% | 23.7% |
| A2 | The gen AI suggestions are helpful to my work. | 4.3% | 6.5% | 26.9% | 25.8% | 36.6% |
| A3 | The gen AI suggestions help me improve my service speed. | 5.4% | 3.2% | 23.7% | 24.7% | 43.0% |
| A4 | The gen AI suggestions help me improve my service quality. | 5.4% | 10.8% | 22.6% | 26.9% | 34.4% |
| A5 | The gen AI suggestions help me find solutions more quickly. | 4.3% | 3.2% | 26.9% | 25.8% | 39.8% |
| A6 | The gen AI suggestions help me understand customer cases more effectively. | 2.2% | 3.2% | 19.4% | 33.3% | 41.9% |
| A7 | The gen AI suggestions allow me to focus more on the customer experience. | 6.5% | 5.4% | 18.3% | 28.0% | 41.9% |
| A8 | The gen AI suggestions help me respond to customers faster. | 3.2% | 2.2% | 17.2% | 26.9% | 50.5% |
| A9 | The gen AI suggestions help me communicate and interact better with customers. | 5.4% | 5.4% | 22.6% | 24.7% | 41.9% |
| A10 | The gen AI suggestions make my work easier. | 6.5% | 4.3% | 22.6% | 24.7% | 41.9% |
| A11 | The gen AI suggestions make me more motivated to participate in work. | 8.6% | 3.2% | 26.9% | 17.2% | 44.1% |

***Note:*** This table shows the distribution of answers from 93 high performers. Respondents rated each statement on a 5-point Likert scale (1 = Strongly Disagree, 5 = Strongly Agree).

## Table A18. Survey Responses of Low-Performing Agents on Their Experience with the Gen AI Assistant

| Do you agree with the following statements? Please indicate how much you agree or disagree with the following statements.<br>No. | Statement | Strongly Disagree | Disagree | Neutral | Agree | Strongly Agree |
|---|---|---|---|---|---|---|
| A1 | The quality of suggestions generated by the gen AI assistant is good. | 2.6% | 4.2% | 29.3% | 35.6% | 28.3% |
| A2 | The gen AI suggestions are helpful to my work. | 2.1% | 3.1% | 23.6% | 32.5% | 38.7% |
| A3 | The gen AI suggestions help me improve my service speed. | 1.6% | 3.1% | 17.8% | 27.7% | 49.7% |
| A4 | The gen AI suggestions help me improve my service quality. | 2.1% | 8.4% | 23.0% | 29.3% | 37.2% |
| A5 | The gen AI suggestions help me find solutions more quickly. | 3.1% | 7.3% | 19.4% | 27.7% | 42.4% |
| A6 | The gen AI suggestions help me understand customer cases more effectively. | 1.6% | 4.7% | 17.8% | 30.9% | 45.0% |
| A7 | The gen AI suggestions allow me to focus more on the customer experience. | 3.1% | 4.7% | 18.8% | 29.3% | 44.0% |
| A8 | The gen AI suggestions help me respond to customers faster. | 1.6% | 3.7% | 16.2% | 27.7% | 50.8% |
| A9 | The gen AI suggestions help me communicate and interact better with customers. | 2.1% | 6.8% | 21.5% | 28.3% | 41.4% |
| A10 | The gen AI suggestions make my work easier. | 3.7% | 6.8% | 23.6% | 24.6% | 41.4% |
| A11 | The gen AI suggestions make me more motivated to participate in work. | 2.1% | 8.9% | 18.3% | 28.3% | 42.4% |

***Note:*** This table shows the distribution of answers from 191 low performers. Respondents rated each statement on a 5-point Likert scale (1 = Strongly Disagree, 5 = Strongly Agree).

## Table A19. Self-Evaluation Scores by Agent Performance Rank

| Do you agree with the following statements? Please indicate how much you agree or disagree with the following statements. | | Mean Score | | | | |
|---|---|---|---|---|---|---|
| No. | Statement | Level 1 | Level 2 | Level 3 | Level 4 | Level 5 |
| B1 | I can easily understand what customers really mean during service interactions. | 4.15 | 4.22 | 4.34 | 4.36 | 4.45 |
| B2 | I can easily relate to and understand how customers feel. | 4.55 | 4.35 | 4.43 | 4.47 | 4.51 |
| B3 | I can quickly review and extract key customer information from the working interface. | 4.20 | 4.17 | 4.30 | 4.40 | 4.66 |
| B4 | I can quickly identify the customer's issue. | 4.25 | 4.13 | 4.38 | 4.43 | 4.67 |
| B5 | Once I identify a customer's issue, I can quickly find an appropriate solution. | 4.25 | 4.08 | 4.29 | 4.42 | 4.51 |
| B6 | I can complete my service tasks efficiently and with ease. | 4.10 | 3.98 | 4.14 | 4.13 | 4.47 |

*Note:* Respondents rated each statement on a 5-point Likert scale (1 = Strongly Disagree, 5 = Strongly Agree).

## Appendix F. Additional Evidence on Agents' Gen AI Usage Decisions

### Table A20. Gen AI Adherence Conditional on Issue Identification Consistency Across Worker Performance Quintiles

| | if adhere to gen AI suggestion | |
|---|---|---|
| | consistent with the identified issue | inconsistent with the identified issue |
| Q1 Agents | 0.404 | 0.370 |
| Q2 Agents | 0.267 | 0.226 |
| Q3 Agents | 0.231 | 0.197 |
| Q4 Agents | 0.191 | 0.160 |
| Q5 Agents | 0.197 | 0.155 |
| All Agents | 0.228 | 0.191 |

### Table A21. Decision Time of Treated Agents: Regression Results

| | decision time |
|---|---|
| | (1) |
| Q2 Agents | 0.053 |
| | (0.078) |
| Q3 Agents | -0.143 |
| | (0.091) |
| Q4 Agents | -0.117 |
| | (0.086) |
| Q5 Agents | -0.035 |
| | (0.093) |
| Day Fixed Effects | Y |
| Time-Variant Controls | Y |
| Observations | 73,659 |
| R-squared | 0.044 |

***Notes.*** The analysis is conducted at the chat level. The sample consists of chats handled by treated agents during the treatment period. *p<0.1; **p<0.05; ***p<0.01. Robust standard errors are in parentheses.

### Table A22. Decision Time and Mouse-Click Behaviors of Treated Agents: Descriptive Statistics

| | | Q1 | Q2 | Q3 | Q4 | Q5 |
|---|---|---|---|---|---|---|
| full sample | decision time (seconds) | 7.705 | 7.415 | 6.121 | 6.337 | 6.408 |
| | # of clicks | 2.341 | 2.701 | 2.412 | 2.422 | 2.437 |